\documentclass[gc, manuscript]{copernicus}

\begin{document}
\nolinenumbers

\title{The Gannon Storm: Citizen science observations during the geomagnetic superstorm of 10 May 2024}


\Author[1,2][maxime.grandin@fmi.fi]{Maxime}{Grandin} 
\Author[3,4]{Emma}{Bruus}
\Author[5,6]{Vincent E.}{Ledvina}
\Author[7]{Noora}{Partamies}
\Author[8,9]{Mathieu}{Barthelemy}
\Author[10]{Carlos}{Martinis}
\Author[11]{Rowan}{Dayton-Oxland}
\Author[12,13]{Bea}{Gallardo-Lacourt}
\Author[14]{Yukitoshi}{Nishimura}
\Author[7]{Katie}{Herlingshaw}
\Author[3]{Neethal}{Thomas}
\Author[4]{Eero}{Karvinen}
\Author[6,15]{Donna}{Lach}
\Author[16]{Marjan}{Spijkers}
\Author[17]{Calle}{Bergstrand}

\affil[1]{Department of Physics, University of Helsinki, Helsinki, Finland}
\affil[2]{Finnish Meteorological Institute, Helsinki, Finland}
\affil[3]{Sodankylä Geophysical Observatory, University of Oulu, Oulu, Finland}
\affil[4]{Taivaanvahti/Skywarden Observation Service, Ursa Astronomical Association, Helsinki, Finland}
\affil[5]{Geophysical Institute, University of Alaska Fairbanks, Fairbanks, AK, USA}
\affil[6]{Aurorasaurus, New Mexico Consortium, Los Alamos, NM, USA}
\affil[7]{University Centre in Svalbard, Longyearbyen, Norway}
\affil[8]{Université Grenoble Alpes, CNRS, IPAG, 38000 Grenoble, France}
\affil[9]{Université Grenoble Alpes, CSUG, 38000 Grenoble, France}
\affil[10]{Center for Space Physics, Astronomy Department, Boston University, Boston, USA}
\affil[11]{University of Southampton, Southampton, United Kingdom}
\affil[12]{Goddard Space Flight Center, National Aeronautics and Space Administration, Greenbelt, MD, USA}
\affil[13]{Department of Physics, The Catholic University of America, Washington, DC, USA}
\affil[14]{Department of Electrical and Computer Engineering and Center for Space Physics, Boston University, Boston, USA}
\affil[15]{Citizen Scientist, Canada}
\affil[16]{Citizen Scientist, The Netherlands}
\affil[17]{Citizen Scientist, Sweden}




\runningtitle{GANNON STORM CITIZEN SCIENCE OBSERVATIONS}

\runningauthor{GRANDIN et al.}

\received{}
\pubdiscuss{} 
\revised{}
\accepted{}
\published{}


\firstpage{1}

\maketitle

\begin{abstract}
The 10~May 2024 geomagnetic storm, referred to as the Gannon Storm in this paper, was one of the most extreme to have occurred in over 20~years. In the era of smartphones and social media, millions of people from all around the world were alerted to the possibility of exceptional auroral displays. Hence, many people not only witnessed but also photographed the aurora during this event. These citizen science observations, although not from scientific instruments operated by observatories or research groups, can prove invaluable in obtaining data to characterise this extraordinary event. In particular, many observers saw and photographed the aurora at mid-latitudes, where ground-based instruments targeting auroral studies are sparse or absent. Moreover, the proximity of the event to the northern hemisphere summer solstice meant that many optical instruments were not in operation due to the lack of suitably dark conditions.
We created an online survey and circulated it within networks of aurora photographers to collect observations of the aurora and disruptions in technological systems that were experienced during this superstorm. We obtained 696~citizen science reports from over 30~countries, containing information such as the time and location of aurora sightings, observed colours and auroral forms, as well as geolocalisation, network, and power disruptions noticed during the geomagnetic storm. We supplemented the obtained dataset with 186~auroral observations logged in the Skywarden catalogue (\url{https://taivaanvahti.fi}) by citizen scientists.
The main findings enabled by the data collected through these reports are that the aurora was widely seen from locations at geomagnetic latitudes ranging between 30$^\circ$ and 60$^\circ$, with a few reports from even lower latitudes. This was significantly further equatorward than predicted by auroral oval models. The reported auroral emission colours, predominantly red and pink and intense enough to reach naked-eye visibility, suggest that the auroral electron precipitation contained large fluxes of low-energy (<\,1~keV) particles. 
This study also reveals the limitations of citizen science data collection via a rudimentary online form. We discuss possible solutions to enable more detailed and quantitative studies of extreme geomagnetic events with citizen science in the future.
\end{abstract}


\introduction[Introduction]  

The aurora has captivated human beings since ancient times, with some of the earliest reports found in an ancient Chinese annal known as the Bamboo Annals and mentioning auroral observations from near Xi'an dating from either $977 \pm 1$~BCE or $957 \pm 1$~BCE \citep{Usoskin2023}. Typically visible at high latitudes around Earth's poles, the aurora can be seen further equatorward during geomagnetic storms \citep[e.g.][]{blake2021}. Such was the case during the extreme geomagnetic storm starting on 10~May 2024, which was the most intense geomagnetic storm since the so-called Halloween storms of October--November 2003, over 20~years ago \citep{greshko2024}. Two names have been put forward to refer to the 10~May 2024 superstorm: ``Mother's Day Storm'' (as it occurred over Mother’s Day weekend in some countries), or ``Gannon Storm'' (in memory of Dr.~Jennifer Gannon). In this paper, we will retain the name of ``Gannon Storm'', both to honour Dr.~Gannon's memory and to acknowledge the fact that Mother's Day does not have a universal date across countries. The multiple coronal mass ejections (CMEs) from the Sun, which led to this geomagnetic storm, originated from a large and complex cluster of sunspots sixteen times the diameter of Earth \citep{spogli2024,kwak2024}. The plasma parameters and interplanetary magnetic field associated with these interacting CMEs produced exceptional driving conditions when encountering the Earth's magnetosphere.

The consequences of extreme geomagnetic storms are not negligible. Human-made space-borne and ground-based infrastructure can be significantly affected. \citet{Hapgood2021} give an interesting overview of the potential space weather effects from such extreme events, focusing on British infrastructure. These include threats to power grid systems posed by geomagnetically induced currents, impacts on satellite communication and global navigation satellite systems (GNSS), blackouts and anomalous propagation of high-frequency radio signals, damage on the onboard electronics of satellites, increased atmospheric drag for low-Earth-orbiting satellites, as well as effects on civil aviation due to enhanced radiation doses in the polar regions. A few studies have undertaken estimations of the potential financial cost of space weather impacts and found values ranging from millions to tens of billion US dollars per day, corresponding to 15--100\% of the daily US GDP, depending on the tested scenario \citep{oughton2017}, or a total of up to a few trillion US dollars in case of a Carrington-level superstorm \citep{eastwood2018}. The September 1859 ``Carrington'' event is one of the largest storms ever documented for which geomagnetic indices have been estimated \citep{siscoe2006}. Much more recently, the 23~July 2012 extreme coronal mass ejection (CME) which was not Earth-directed could have produced a geomagnetic storm on par with -- if not more intense than -- the Carrington Storm \citep{Baker2013}. The aforementioned studies of financial costs of space weather further underline that the obtained estimates bear large uncertainties, as evaluating the total costs for economy increasingly relying on technology is extremely challenging.

When extreme geomagnetic storms occur, the auroral ovals extend equatorward, usually beyond the regions where most of the ground-based scientific instruments (optical and radar) designed for space physics studies are located \citep[e.g.][]{johnsen2013,kataoka2024}. Therefore, the most severe events are likely not to be properly captured by our routine science observations apart from the few instruments at mid-latitudes and satellite measurements which inherently provide sparse coverage. For instance, the Super Dual Auroral Radar Network \citep[SuperDARN][]{Nishitani2019} mid-latitude radars provide an uneven coverage across longitudes, with large gaps in the European, Russian and southern mid-latitude sectors. Superstorms are also poorly described by numerical models, as the driving conditions lie beyond the regime of their validity. For instance, the OVATION Prime model of the auroral oval \citep{newell2014} has been fitted for driving conditions up to those corresponding to Kp index values of 8+. During the Gannon Storm, the Kp index reached the value of 9 during three three-hour periods \citep{spogli2024}. As extreme conditions are being largely absent from training datasets, AI-based models are currently limited in making realistic forecasts of geomagnetic conditions during superstorms \citep[e.g.][]{Oliveira2021}.

For these reasons, finding other data sources to study the most extreme geomagnetic storms such as the Gannon Storm is necessary. With the recent tremendous improvement of commercially available off-the-shelf (COTS) and smartphone camera systems, large numbers of people now have the capability to take relatively high-quality photos of the aurora and night sky. Using such images as data for scientific studies has proved successful in a growing number of studies which showcase examples of citizen science applied to space physics. Often, those studies have made use of selected photographs of the aurora taken by citizen scientists to investigate elusive optical phenomena for which observations by scientific instruments were either missing or not sufficient. For instance, the phenomenon known as STEVE (a backronym for Strong Thermal Emission Velocity Enhancement) was first uncovered thanks to citizen scientists \citep{macdonald2018} and is often accompanied by vertical green structures dubbed the picket fence. STEVE and the picket fence have given rise to numerous publications relying on citizen science imagery \citep[e.g.][]{archer2019,Mende2019,Martinis2021,nishimura2023}. Furthermore, the relationship between STEVE and stable auroral red (SAR) arcs has been examined thanks to citizen science \citep{martinis2022}. Emission structures of very small spatial scales have also been investigated involving citizen scientists, either using photographs they took \citep[e.g. the so-called ``streaks'' occasionally seen along with the picket fence;][]{semeter2020} or having them participate in auroral form classification \citep[e.g. fragmented aurora-like emissions;][]{whiter2021}. Finally, ``dunes'', which exhibit wave-like patterns in the diffuse green aurora, have also been investigated largely relying on citizen science data from auroral photographers \citep{palmroth2019,grandin2021,he2023}.

Citizen science applied to space physics has the potential to produce scientifically valuable data not only for the study of specific phenomena, as discussed above, but also by providing an agile way to collect observations on a more general level \citep{ledvina2023}. Collaborative aurora-sighting reporting platforms such as Aurorasaurus \citep{macdonald2015} have proved very powerful in providing ground-truth validation of auroral boundaries derived from empirical models \citep{cosar2018_boundaries}. Aurorasaurus has also demonstrated its efficacy in generating auroral visibility alerts in real time \citep{case2016_alerts}. Moreover, the submitted auroral visibility reports can be accompanied by images which are geotagged and made available for scientific use after validation \citep{cosar2018_data}. Another collaborative catalog of observations providing invaluable data for auroral studies is Skywarden (\url{https://taivaanvahti.fi} from its original Finnish name: Taivaanvahti; see Sect.~\ref{sec:datacollection}). This aurora observation platform established in 2011 \citep{ursa2021} contains more than 10 000 open-access, content-verified auroral observations accompanied with one or several photographs. Skywarden's observations have been used in relation to the ``dunes'' \citep{palmroth2019}, in the discovery of proton injection initiated red arc with green diffuse aurora \citep{nishimura2022} and in research of other atmospheric and astronomical phenomena \citep{gritsevich2014,moilanen2022}.

The Gannon Storm was the first geomagnetic event of such magnitude occurring during the era of social media and COTS cameras and smartphones capable of photographing the aurora. Therefore, it is an unprecedented opportunity to investigate to what extent citizen scientists can provide observations that could be used to improve our characterisation of processes occurring during extreme geomagnetic storms. It is also a unique opportunity to estimate how many citizen scientists from all around the world can be mobilised to provide reports on what they saw during the event, including among those who had no prior experience with auroral observations and were not aware of projects like Skywarden or Aurorasaurus. To test these ideas, the ARCTICS (Auroral Research Coordination: Towards Internationalised Citizen Science, \url{https://collab.issibern.ch/arctics/}) working group supported by the International Space Science Institute in Bern, Switzerland, conducted an online survey which was distributed as widely as possible among our network of collaborators and among online networks of aurora chasers. This survey enabled citizen scientists to report whether they saw the aurora or experienced disruptions in technological systems during the superstorm (see Sect.~\ref{sec:datacollection}).

The objective of this paper is to evaluate how much data can be collected thanks to citizen science during extreme geomagnetic events and what kind of scientifically usable information can be obtained from citizen scientist observations. We will pay particular attention to scientific results that would prove difficult to obtain relying solely on existing scientific instruments and models. Section~\ref{sec:context} presents the geophysical context of the Gannon Storm. Section~\ref{sec:datacollection} describes the collection and pre-processing of citizen science observations of the aurora and technological disruptions during the storm, and in Sect.~\ref{sec:analysis} we analyse the citizen science data and examine what scientific conclusions on the storm can be inferred from this dataset. We discuss the limitations and challenges of the collected citizen science observations and propose solutions to overcome them in Sect.~\ref{sec:discussion}, and we summarise our findings in Sect.~\ref{sec:conclusions}.

\section{Geophysical context}
\label{sec:context}

\begin{figure}[t]
  \includegraphics[width=.6\textwidth]{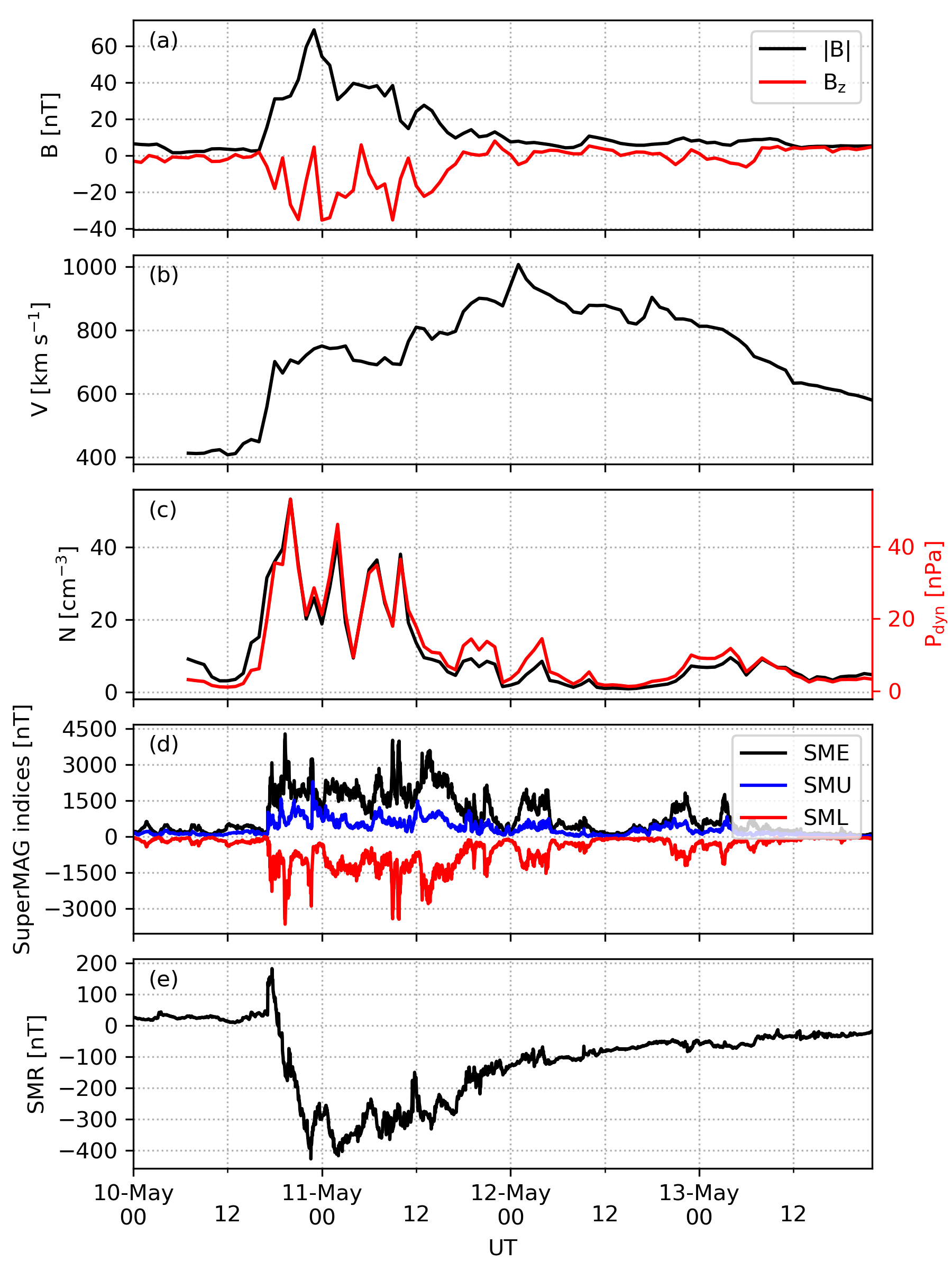}
  \caption{Driving and geomagnetic conditions during the Gannon Storm. (a)~Interplanetary magnetic field magnitude (black) and $B_z$ component (red). (b)~Solar wind speed. (c)~Solar wind density (black) and dynamic pressure (red). (d)~SME (black), SMU (blue) and SML (red) indices. (e)~SMR index.}
  \label{fig:driving}
\end{figure}

Between 8 and 10~May 2024, an Earth-facing sunspot region AR 3664 produced multiple X-class flares associated with halo CME, indicating their propagation towards the Earth. These CMEs may have collided and merged, resulting in a complex magnetic cloud structure. When this CME structure reached the Earth, it triggered a geomagnetic storm categorised as ``extreme'' (G5-class according to the NOAA Space Weather Scales, \url{https://www.swpc.noaa.gov/noaa-scales-explanation}). A detailed analysis of the driving conditions and of the global geomagnetic response they led to can be found in \citet{kwak2024}.

We retrieved the provisional 1-hour OMNI data \citep{OMNI_data_1hr} to visualise the key interplanetary magnetic field (IMF) and solar wind parameters during 10--13~May 2024. Those data have been measured by various spacecraft orbiting around the L1 Lagrange point of the Sun--Earth system, and propagated to the terrestrial bow shock. Figure\,\ref{fig:driving} shows the time series of the solar wind and IMF parameters during the event. The IMF magnitude (black line) and $B_z$ (north--south) component (red line) are presented in Fig.\,\ref{fig:driving}a. We can see that around 16~UT on 10~May 2024, the IMF magnitude sharply increases from a few nanoteslas to the maximum of 69~nT at 23~UT. It remains significantly elevated throughout 11~May, and does not decrease below 10~nT until the early hours of 12~May. During most of this time, $B_z$ is predominantly southward (negative), reaching values as low as --35~nT at 21~UT on 10~May, 0~UT on 11~May, and 9~UT on 11~May. This strongly southward IMF component produces sustained magnetic reconnection at the dayside magnetopause, fuelling plasma convection in the inner magnetosphere and in the polar cap and subsequently leading to reconnection in the magnetotail associated with the injection of charged particles into the magnetosphere.

Figure\,\ref{fig:driving}b shows the solar wind speed. The data gap in the earliest hours of the considered time period is related to missing measurements. It can be seen that at the same time as the IMF compression started, the solar wind speed increases from about 400~km\,s$^{-1}$ to over 700~km\,s$^{-1}$ at the time of peak IMF field strength. The speed continues to increase until reaching the maximum of 1006~km\,s$^{-1}$ at 1~UT on 12~May, after which it gradually decays. It is still of the order of 600~km\,s$^{-1}$ in the late hours of 13~May.

The solar wind density and dynamic pressure strongly increase in tandem on 10~May, reaching the maximum of 53~cm$^{-3}$ and 53~nPa at 20~UT, respectively (Fig.\,\ref{fig:driving}c). The decay of these two parameters happens by noon on 11~May, which is in agreement with the decay of the IMF structure and the strongest magnetic variability (SME index, see below), but much faster than that of the solar wind speed, which remains high past 13~May.

To assess the geomagnetic response to these extreme driving conditions, we consider the provisional SuperMAG indices, derived from ground-based magnetometer measurements \citep{SuperMAG_technique}. These indices have been retrieved at 1~minute resolution. Figure\,\ref{fig:driving}d shows the SuperMAG electrojet (SME) index, together with its upper and lower envelope components, SMU and SML \citep{SuperMAG_electrojet}. These indices provide a measure of the auroral electrojet intensity and are therefore used as a proxy for substorm activity. It is clear that an abrupt increase in substorm activity takes place shortly after the arrival of the solar wind disturbances: SME increases from under 200~nT to over 800~nT in the span of two minutes (between 17:05~UT and 17:07~UT on 10~May). SME continues to increase rapidly and reaches an initial peak of 3077~nT at 17:38~UT, and a second peak of 4276~nT at 19:19~UT. Substorm activity remains extreme (i.e. SME\,>\,2000~nT for a majority of the time) until $\sim$18~UT on 11~May, and elevated until about 6~UT on 12~May. We can see that, throughout the disturbed period, both SMU and SML exhibit elevated values (in absolute units), indicating recurring intensification in the westward and eastward auroral electrojets, respectively. To put this in perspective, we looked at the SME index values since the Halloween storms of 2003. We found that SME exceeded 3000~nT during only 21~geomagnetic events between 20~November 2003 (excluded) and 31~December 2023. It exceeded 4000~nT only 7~times during that 20-year period (and only twice during the previous solar cycle [SC24: 2008--2019]: on 28~May 2011 and on 8~September 2017).

Finally, Fig.\,\ref{fig:driving}e shows the SuperMAG ring current (SMR) index \citep{SuperMAG_ringcurrent}, which provides a measure of the ring current intensity and is hence used as a proxy for geomagnetic storm intensity. A very prominent storm sudden commencement \cite[SSC;][]{akasofu2005} signature is visible, starting at 17:07~UT on 10~May and reaching a peak of 182~nT at 17:39~UT. Those two times are consistent with the solar wind dynamic pressure starting to increase, indicating a strong compression of the dayside magnetopause due to the arrival of the CME(s) at Earth. After the SSC signature, SMR decreases rapidly until reaching a first minimum value of $-$427~nT at 22:36~UT on 10~May. A second minimum is attained around two hours later ($-$417~nT at 2:06~UT on 11~May) during the geomagnetic storm's main phase. The recovery phase is not fully over by the end of 13~May, as SMR values are still of the order of $-$20~nT. To put this in perspective, SMR went below $-$200~nT only 8~times between 20~November 2003 (excluded) and 31~December 2023. It went below $-$300~nT only once (on 8~November 2004, meaning that this never happened during SC24).

\section{Citizen science data collection and pre-processing}
\label{sec:datacollection}

The Gannon Storm produced spectacular auroral displays that reached exceptionally low latitudes. This offered the opportunity for many people to witness and often photograph the northern and southern lights for the first time in their life. These observations bear the potential to be used as data to study this extreme event, supplementing optical data from scientific instruments which are mainly deployed in the polar regions \citep[see e.g. Fig.\,1 in][]{alfonsi2022}.

To evaluate the potential of citizen scientist observations in providing insightful information on the Gannon Storm and its effects, we set up an online form enabling citizen scientists to report on their observations of auroral displays and technological disruptions. This survey was hosted on Google Forms and distributed as widely as possible across online aurora chasing communities. The form accepted answers between 18~May and 24~June 2024. The list of questions the survey contained is provided in Appendix~\ref{sec:surveyquestions}. Besides ours, similar initiatives have arisen from other teams who set up surveys to collect citizen science observations during the Gannon Storm and whose analysis is underway \citep[e.g.][]{Hayakawa2024}.

By the end of the data collection period, a total of 696~reports had been submitted. Given that the data collection platform (Google Forms) did not enable imposing strong formatting constraints in the response fields, the data had to be pre-processed before carrying out the analysis. The pre-processing consisted of the following:
\begin{enumerate}
    \item Harmonising all time zone information with respect to the Universal Time Coordinate (UTC) under the format ``UTC$\pm$??'' (e.g. ``UTC+00'', ``UTC+12'', ``UTC-06'', etc). Many responses referred to time zones either by their name (e.g. ``Mountain Daylight Time'') or by the corresponding country/city (e.g. ``Paris time zone''). Among the respondents, 109 indicated their observation time directly as UTC.
    \item Harmonising all geolocalisation information under the format ``latitude, longitude'' with 0.01$^\circ$ precision (e.g. \mbox{``48.12, -1.36''}; \mbox{``-36.84, 174.73''}; etc.). We retained this level of precision, as it is sufficient for our purposes in this study. When no geographic coordinates were provided, we used the ``city, state/region/province, country'' information to search for an approximate geolocation. This was needed for 302~of the reports. In five cases, the provided geographic information was not sufficient (e.g. country only); these observations have hence been excluded from the analysis. The harmonisation to the desired format concerned the other 389~coordinates which were input with higher precision or with the degree, minute, second format.
    \item Correcting the dates which had been input in the American format (MM/DD/YYYY) or had a typo in the year or month, when the intended value was obvious. A total of 44~dates had to be corrected; in 3~cases, the intended value was unclear, and the corresponding reports have thus been excluded from the analysis.
    \item Correcting the times which had not been entered in the 24-hour format (i.e. which used a 12-hour format, although no ``am'' or ``pm'' information was appended). This ambiguity was easily spotted and corrected, as no location from which the reports were issued could have possibly had visible aurora during daytime. There were 32~such corrections needed.
\end{enumerate}

The pre-processed data used in this study are published as an openly available dataset \citep{survey_data}. Note that in order to protect the privacy of the survey participants, we downgraded the precision in their geolocalisation to 0.1$^\circ$ accuracy in this dataset. Moreover, the observations are not tied to the observers' names, but the citizen scientists who wished to be acknowledged by name in this paper are listed alphabetically in Appendix~\ref{sec:csnames}.

In addition to the material collected via the survey form, observations during the storm period were extracted through Skywarden's public data interface. Skywarden is a public, content-moderated observation system established by the Ursa Astronomical Association \citep{ursa2021}. Through its public Application Programming Interface \citep[API;][]{bruus2024}, the data collected in Skywarden can be extracted in either HTML, CSV, XML or JSON formats. This additional dataset from Skywarden contains 186~observations. Skywarden uses an online-form-based interface for gathering observations of different astronomical and atmospheric phenomena, including the aurora. The interface guides users to provide observation data directly in a defined and structured format.

Skywarden uses the ISO 8601 basic format for storing date and time information. Conversions from local to UTC time are done automatically by using Google's Time Zone API. In case user-set coordinate information is missing from an observation, Google's Geolocation API is used for retrieving the missing WGS84-format coordinates. Skywarden provides information of visually observed aurora colours and special aurora structures like SAR arcs, STEVEs, picket fences, dunes, etc. The identifications are checked by the moderator team before the material is published on the website. Because of the pre-existing mechanisms for securing adequate data quality, no additional modifications were made to superstorm-event citizen science data extracted from Skywarden.

\section{Citizen science data analysis}
\label{sec:analysis}

\subsection{Auroral observations}
\subsubsection{Overview of the reports}

\begin{figure}[t]
  \includegraphics[width=\textwidth]{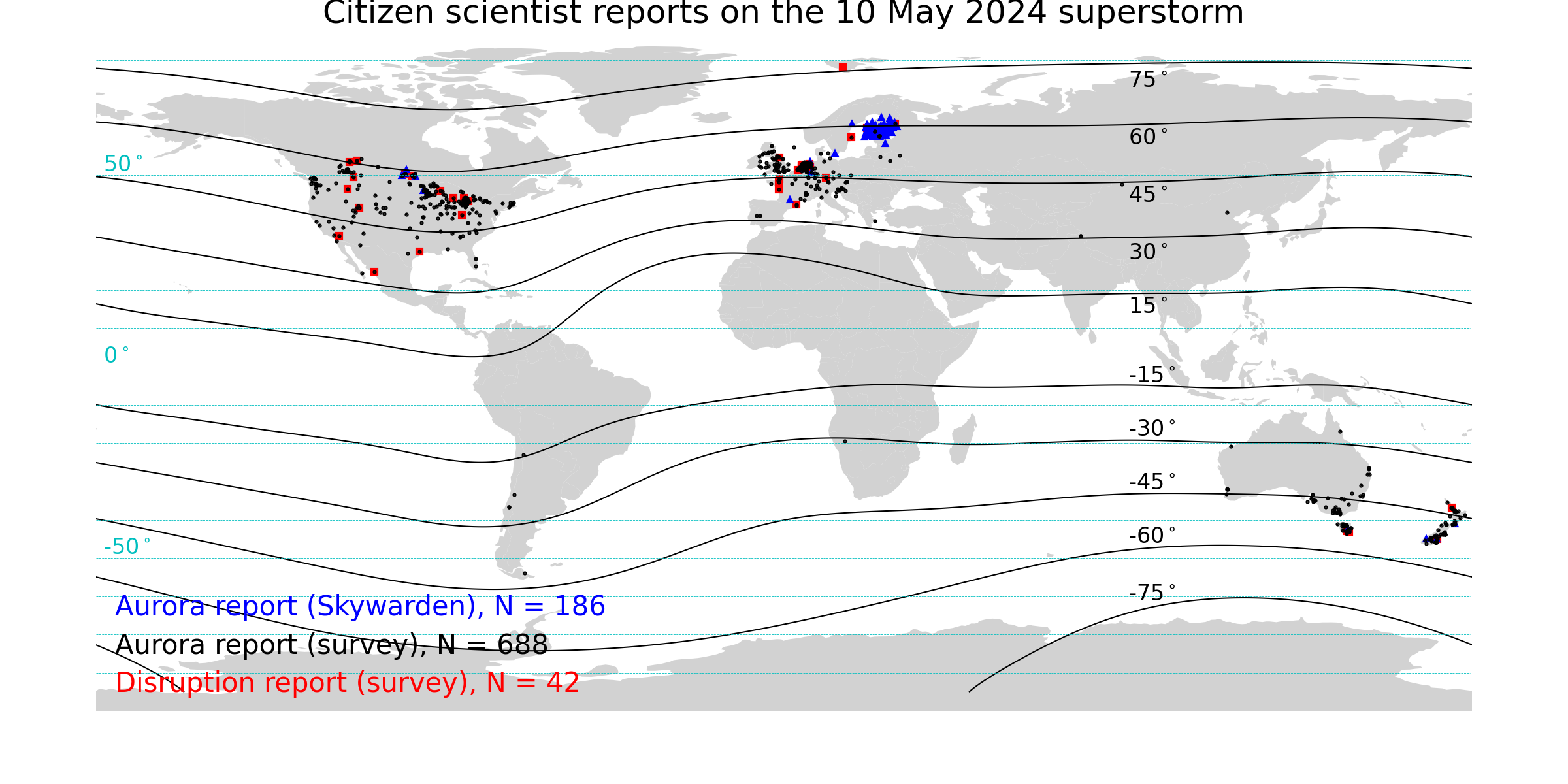}
  \caption{Map of the received reports on aurora sightings from Skywarden (blue dots) and our survey (black dots), and experienced disruptions from the survey (red squares) during the Gannon Storm. The black isocontour lines indicate selected geomagnetic latitudes, and the thin cyan lines show the geographic latitude grid.}
  \label{fig:all_reports}
\end{figure}

Figure\,\ref{fig:all_reports} presents the locations from which citizen scientists reported aurora sightings and/or experienced disruptions during the Gannon Storm. On the map, each black dot corresponds to an aurora sighting reported via the online survey. A total of 688~citizen scientists (out of the 696~respondents) indicated that they saw and/or photographed the aurora. The 42~red squares show where technological disruptions were reported via the online form. The blue triangles correspond to auroral observations logged in Skywarden; there are 186 of these. Most (170) of these latter observations are from Finland, but it is worth noting that the Skywarden data also include observations from Canada (4), New Zealand (3), the Netherlands (2), Sweden (2), Australia (1), Estonia (1), France (1), Germany (1), and the United States (1).

From the map, it can be seen that most reports come from three regions: Europe, North America, and Oceania. Only ten reports come from other regions of the world: four from Chile, two from India, two from China, one from Namibia, and one from Argentina. This uneven coverage of the emerged landmass can be explained by several factors:
\begin{enumerate}
    \item Collaborator network: The survey was circulated by citizen scientists from our network of collaborators who are mainly able to reach observers from their own country.
    \item Language: The survey was set up in English, making it less accessible to observers from non-English-speaking countries.
    \item Population density: Some regions where few to no data points are available are sparsely populated (e.g. Patagonia, Russia, Central Australia, Antarctica\dots).
    \item Lack of darkness: Since the geomagnetic storm took place less than 1.5~month away from the summer solstice in the northern hemisphere, high-latitude observations of the aurora were impeded by the lack of darkness (e.g. Alaska, northern Canada, Greenland, Iceland, northern Fennoscandia, Svalbard). The northernmost observation of the aurora during that storm reported in Skywarden was made just shy of 65$^\circ$ geographic latitude. In the most northerly observations, the Sun was just roughly 5$^\circ$ below the horizon.
\end{enumerate}
The first two points, in particular, may explain the low number of reports from Eastern Asia, while evidence of aurora sightings is found in news articles from Japan (e.g. \url{https://www3.nhk.or.jp/nhkworld/en/news/backstories/3308/}), China (e.g. \url{https://www.chinadaily.com.cn/a/202405/15/WS664411d1a31082fc043c7279.html}), and Korea \citep[e.g.][]{kwak2024}.

\begin{figure}[t]
  \includegraphics[width=.7\textwidth]{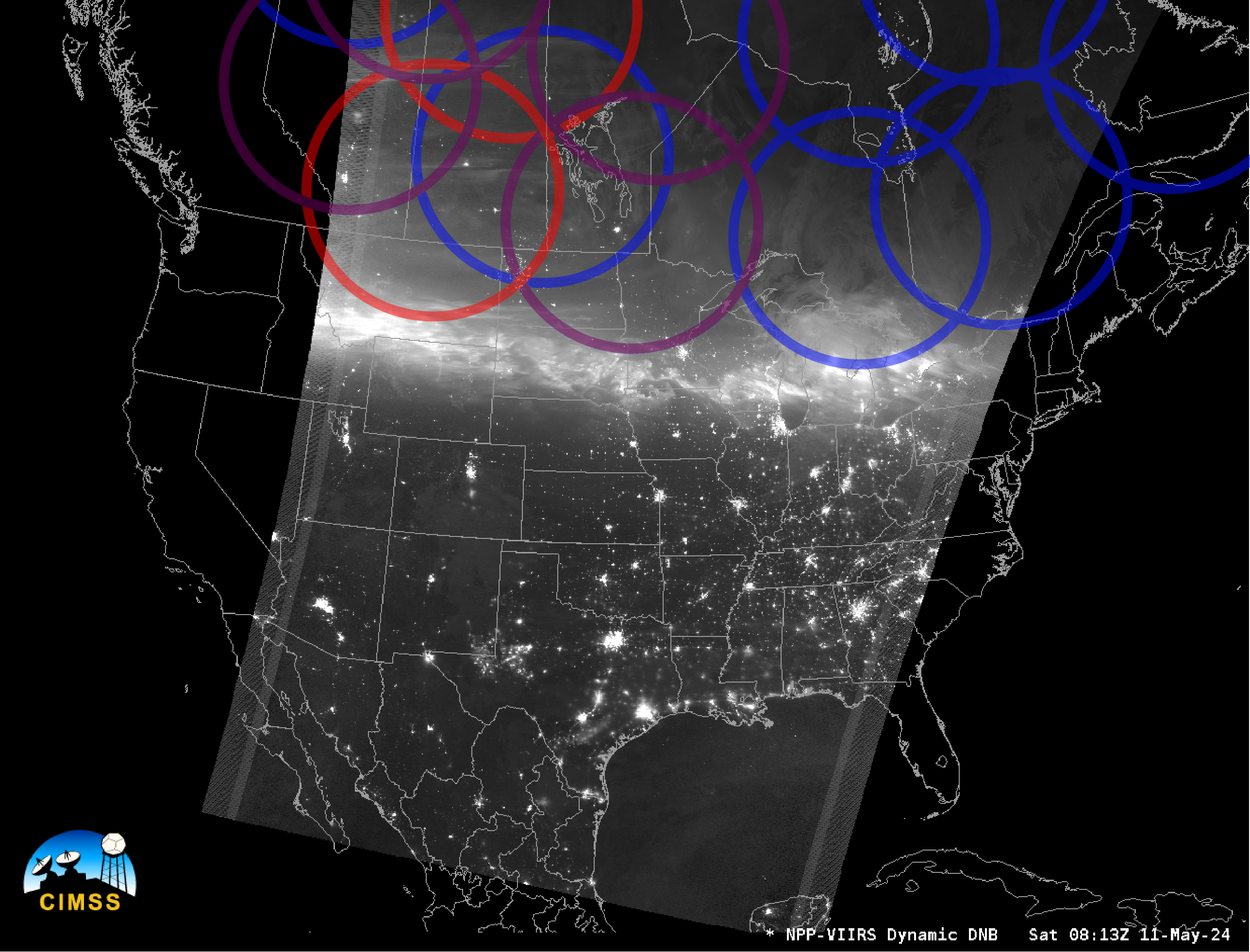}
  \caption{Image from the Visible Infrared Imaging Radiometer Suite (VIIRS) Day Night Band (DNB) onboard the Suomi NPP satellite showing a portion of the auroral over the central United States. Data were captured during the geomagnetic superstorm at 08:13~UT on 11 May 2024. FOVs of THEMIS (blue) and TREx (red) ASIs are overplotted. Co-located TREx and THEMIS ASIs are shown in purple.}
  \label{fig:oval}
\end{figure}

While the limited coverage of reported observations of the aurora during the event does not enable us to obtain a comprehensive picture of the extent of the auroral oval, it is worth pointing out that the citizen scientist reports altogether provide information which is largely missing from scientific measurements. Indeed, most of the reports come from regions situated at mid-latitudes where scientific infrastructure targeting auroral studies (optical instruments, ionospheric radars\dots) is sparse \citep[e.g.][]{alfonsi2022}. Besides, at high latitudes in the northern hemisphere, many of the optical instruments observing auroral emissions were not in operation during the storm since the background sky was not dark enough. This is the case for most optical instruments in the European sector. However, some instruments in North America, such as THEMIS all-sky imager (ASI), TREx, MANGO, and others \citep[e.g.][]{Donovan2006,Liang_2023,bhatt2023}, were able to record parts of the auroral activity.  At some points during the geomagnetic storm, the auroral oval extended equatorward, beyond the field of view (FOV) of the TREx and THEMIS ASI networks (see Fig.\,\ref{fig:oval}). We will see below that scientific conclusions on this geomagnetic storm can be drawn from the citizen science data, even without sophisticated analyses of photographs taken by the citizen scientists.

\subsubsection{Magnetic local time -- geomagnetic latitude distribution of the observations}

\begin{figure}[t]
  \includegraphics[width=\textwidth]{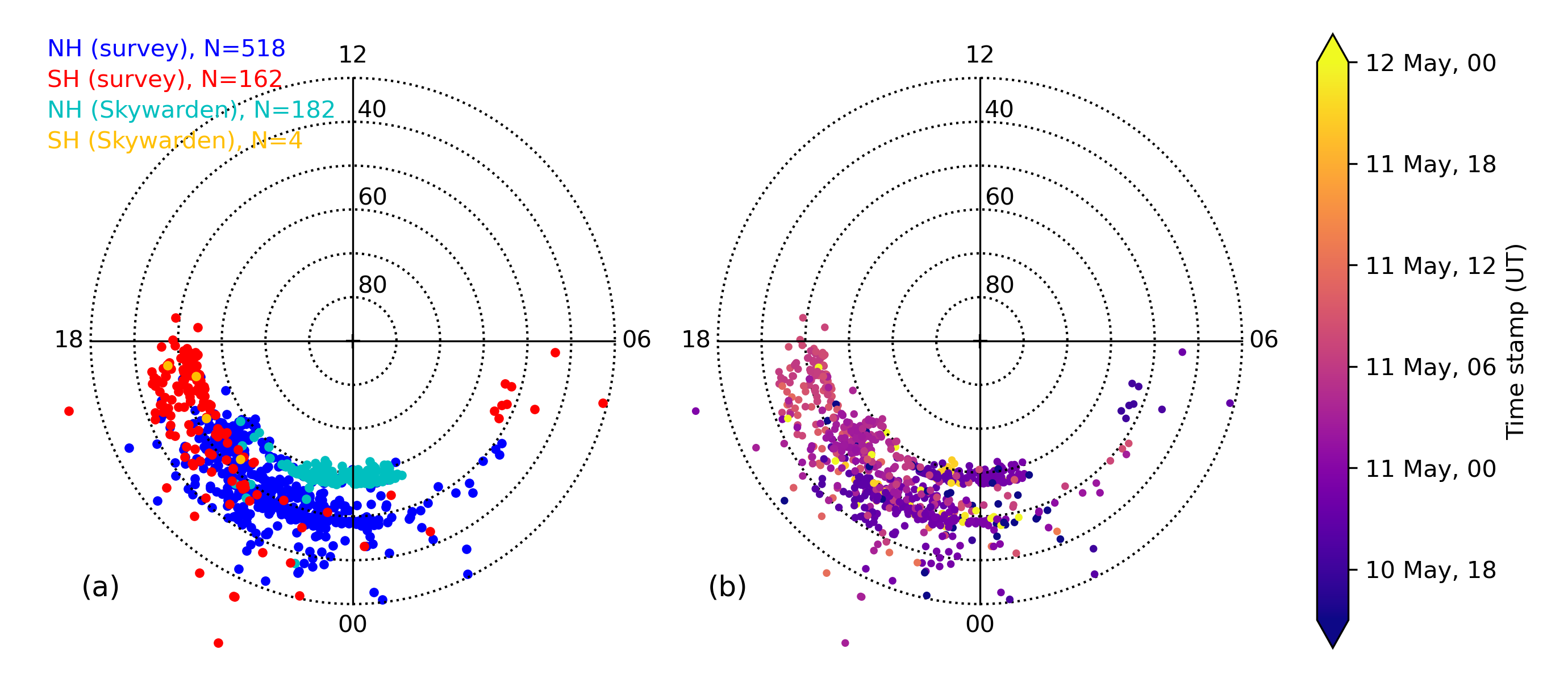}
  \caption{Distribution of aurora sightings during the Gannon Storm as a function of geomagnetic latitude and magnetic local time. The distribution is shown as a function of (a) data source and hemisphere, and (b) observation time stamp. NH: northern hemisphere; SH: southern hemisphere.}
  \label{fig:mlt_mlat}
\end{figure}

A simple, first result that can be obtained by analysing the collected citizen scientist observations is a measure of the auroral oval extent during the Gannon Storm. Figure\,\ref{fig:mlt_mlat} displays the locations from which citizen scientists reported seeing the aurora as a function of geomagnetic latitude (radial coordinate) and magnetic local time (MLT; angular coordinate). Such an approach to assess the auroral oval extent is similar to what projects such as Aurorasaurus \citep{macdonald2015} can routinely provide. The geomagnetic coordinates of observation sites have been calculated using the Python \texttt{aacgmv2} library \citep{aacgmv2,shepherd2014}. Note that observations from the entire duration of the storm and from all regions are shown together in the figure.

In Fig.\,\ref{fig:mlt_mlat}a, the observations are colour-coded to indicate the source (survey or Skywarden) and hemisphere to which they correspond. Clearly, most of the aurora observations took place in the evening sector with the majority of points being confined within 18--00~MLT. Rather than an indication that the auroral activity was more intense in the evening sector than in the morning sector, this asymmetry can be explained by human behaviour -- citizen scientists are more prone to viewing the aurora in the evening hours rather than waking up shortly after midnight to observe it in the morning. However, we note that a non-negligible number of observations were made between 00 and 06~MLT. One can moreover notice that dusk (18--20~MLT) and dawn (04--06~MLT) observations all came from the southern hemisphere; this is because of the near-summer conditions in the northern hemisphere leading to short nights. In terms of geomagnetic latitude, we have most observations spanning from 40$^\circ$ to 60$^\circ$, and a significant number coming from lower latitudes. The most-equatorward report of aurora observation comes from San Pedro de Atacama, Chile, at a geomagnetic latitude of $-14.5^\circ$. However, at such a low latitude, it is likely that the cause for the observed red emissions was an unusually bright airglow display, presumably linked to the extreme geomagnetic activity. Nonetheless, the auroral oval expansion was exceptional during the geomagnetic storm. This is supported by Fig.\,\ref{fig:oval} which shows an extended portion of the auroral oval in the northern hemisphere at 8:13~UT on 11~May 2024, when storm and substorm activity was very high (see Fig.\,\ref{fig:driving}d--e). The aurora was overhead of $\sim$50$^\circ$ geomagnetic latitude at that time.

We can obtain a more detailed view on the latitudinal extent of the auroral oval during the storm by considering Fig.\,\ref{fig:mlt_mlat}b where the data points are colour-coded as a function of their corresponding time stamps, between 18~UT on 10~May 2024 and 12~UT on 12~May. Observations beyond this time window (31~earlier, 11~later) are shown in the figure with the colour corresponding to the boundary of the time interval. From the colour of the points, we can conclude that the most equatorward sightings of the aurora took place during the late hours (in UT) of 10~May 2024, with clusters of dark-blue dots around 40$^\circ$ geomagnetic latitude in the pre-midnight sector and around 50$^\circ$ latitude near the dawn sector. The aurora was also visible at high latitudes (around 60$^\circ$) near midnight, indicating that the auroral activity was spanning across at least 20$^\circ$ in geomagnetic latitude. Purple dots (corresponding to 00--06~UT on 11~May) are mainly confined within 50--60$^\circ$ geomagnetic latitude, consistent with a period of lower (although still above 1500~nT) SME index values (see Fig.\,\ref{fig:driving}d). During the hours spanning around 12~UT on 11~May, we mostly see near-dusk observations from the southern hemisphere, as Australia and New Zealand were the only places with suitably dark conditions. Reports of aurora sighting from as low as 40$^\circ$ geomagnetic latitudes came from this region, as SME peaked above 3000~nT. Finally, during the late hours of 11~May and beyond (yellow dots), the few reports of aurora that were made come from 50--60$^\circ$ geomagnetic latitude, again consistent with subsiding auroral activity as measured by the SME index.

The data shown in Fig.\,\ref{fig:mlt_mlat} do not strictly give the geomagnetic latitude at which auroral emissions occurred. Indeed, without knowing the elevation of the aurora from a given observation place, there is an uncertainty in the geomagnetic latitude at which the auroral emissions are taking place. Assuming minimum elevations of 5$^\circ$ above the horizon for the reported auroral sightings, and considering 250~km (110~km) as the altitude of the red (green) emission, the oval boundary can be inferred with an uncertainty of $\pm$1284~km (743~km) at best, which corresponds to $\pm$11.5$^\circ$ ($\pm$6.7$^\circ$) in geomagnetic latitude. Despite those rather large uncertainties, this is still a valuable quantitative estimate of the equatorward boundary of the auroral oval to supplement the few available scientific-grade observations during extreme events such as the 10~May 2024 geomagnetic storm.

\begin{figure}[t]
  \includegraphics[width=\textwidth]{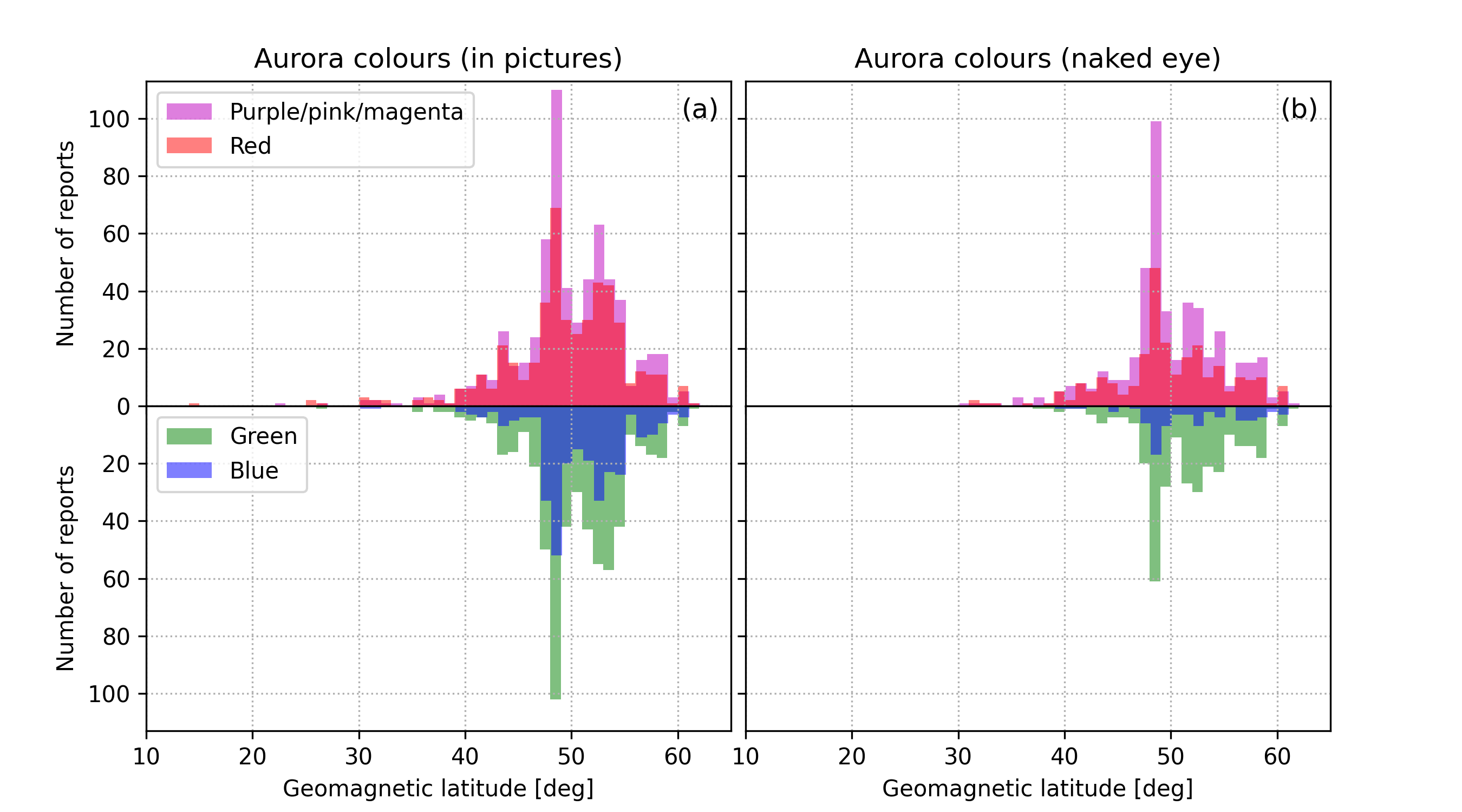}
  \caption{Reported colours of the aurora (a)~in pictures and (b)~visible to the naked eye during the Gannon Storm, as a function of geomagnetic latitude. The data from both hemispheres are blended together.}
  \label{fig:aurora_colour}
\end{figure}

\subsubsection{Reported auroral colours}

Citizen scientists can also report the colours observed in the aurora which is scientifically valuable information. Figure\,\ref{fig:aurora_colour} shows histograms of the reported colours as a function of geomagnetic latitude, both as appearing in pictures (Fig.\,\ref{fig:aurora_colour}a) and as visible to the naked eye (Fig.\,\ref{fig:aurora_colour}b). We consider four auroral emission colours in this study: red, green, blue, and purple/pink/magenta. While the first three typically come from the atomic oxygen 630~nm and 558~nm emissions and the molecular nitrogen ion 428~nm emission, respectively, the purple/pink/magenta colour generally results from a mix of emissions \citep{Sandholt2002} or can be the result of blue emissions seen in twilight conditions (see below).

Comparing the two panels of Fig.\,\ref{fig:aurora_colour}, we can see that the four considered colours were reported by the observers, both in pictures and seen with the naked eye. This suggests that the auroral displays were not only reaching exceptionally low latitudes, but they were also of exceptional brightness, since colours generally cannot be distinguished to the naked eye when the aurora is dim. The human eye's sensitivity to colour maximises at around 530--560~nm \citep{malacara2011}, which makes the visual observations of red colour exceptional \citep{schnapf1987}. A very large number of reports indicated red emissions seen with the naked eye, which implies very bright auroral displays during the geomagnetic storm.

Focusing on the latitudinal distribution of the observed colours, one can see that a wide range of geomagnetic latitudes have reports of all four colours. It is of particular interest to note that green emissions were seen down to about 38$^\circ$ geomagnetic latitude, and photographed down to about 35$^\circ$. This is very unusual, as when the aurora reaches mid-latitudes, it is mainly red emissions at high altitudes that are observed \citep{green2006}. Red (and purple/pink/magenta) aurora was seen and photographed by a few observers below 30$^\circ$ -- down to 14.5$^\circ$, although as mentioned above in that extreme case it was likely strong airglow -- geomagnetic latitude, consistent with the fact that only the uppermost part of the aurora could be seen just above the poleward horizon for those observations, and the upper part red emission can strongly enhance during geomagnetic storms driven by dense solar wind \citep{kataoka2024}. The low latitudes for auroral observations are beyond the NOAA geomagnetic storm category G5 definition, which says that aurora is typically seen down to about 40$^\circ$ geomagnetic latitude (see \url{https://www.weather.gov/media/publications/assessments/SWstorms\_assessment.pdf}).

\begin{figure}[t]
  \includegraphics[width=\textwidth]{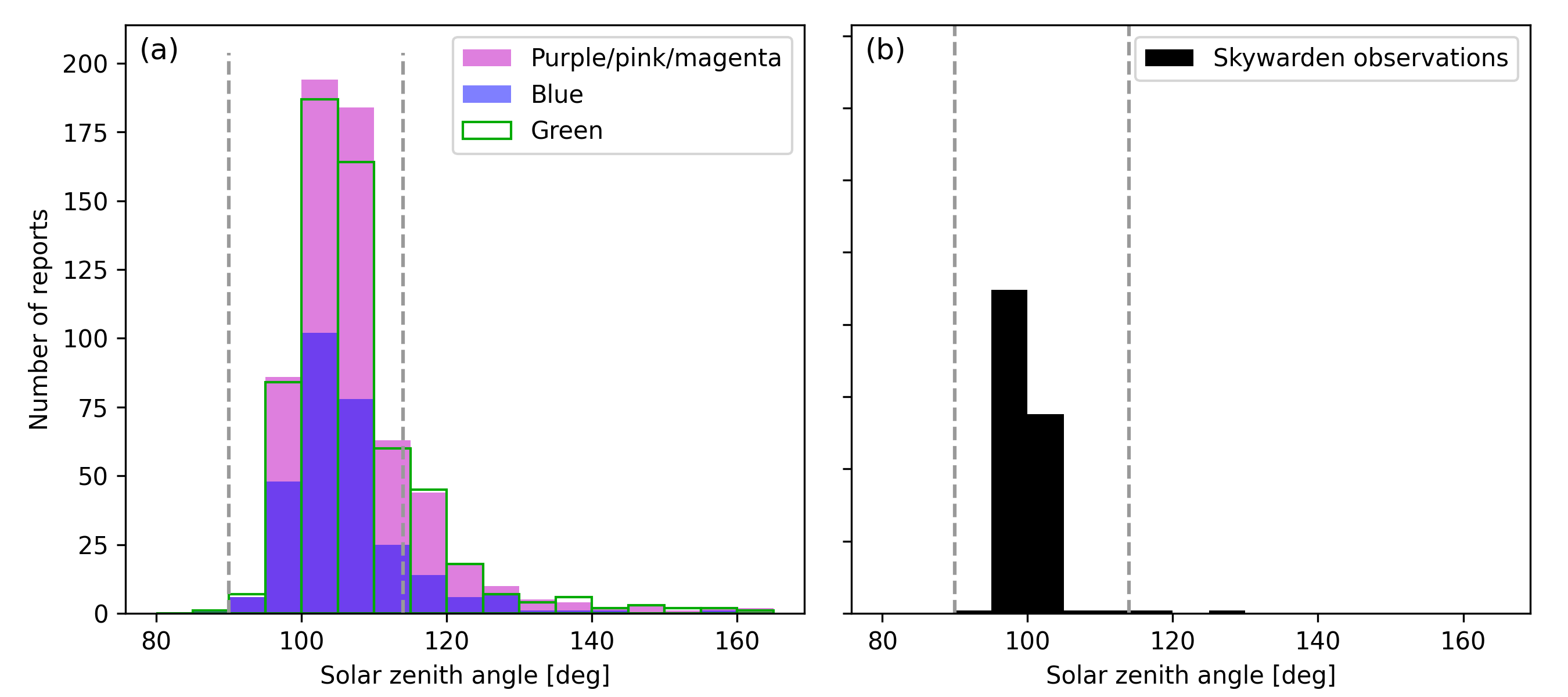}
  \caption{Distribution of solar zenith angle values corresponding to (a)~the observation of purple/pink/magenta, blue, and green auroral emissions in the survey reports, and (b)~in Skywarden observations. The vertical grey dashed lines indicate solar zenith angle values of 90$^\circ$ and 114$^\circ$, corresponding to solar elevations comprised between 0$^\circ$ and $-$24$^\circ$ below the horizon.}
  \label{fig:sza}
\end{figure}

A striking feature visible from Fig.\,\ref{fig:aurora_colour} is that a large number of observers reported red and purple/pink/magenta aurora, both in pictures and as naked-eye observations. While the intense red aurora suggests that large fluxes of low-energy ($\ll$\,1\,keV) electrons depositing their energy above about 200~km were precipitating during the event \citep{fang2010}, the purple/pink/magenta emissions can result from two main mechanisms. First, such colours can be seen in the E-region aurora in presence of \mbox{$\sim$1--10~keV} electron precipitation, when emissions from molecular nitrogen produce pink hues in the aurora \citep{whiter2023}. Second, similar colours can be obtained at high altitudes (above about 150--200~km) via resonant scattering of sunlight by N$_2^+$ excited in presence of low-energy electron precipitation and ion upflow \citep{bates1949,broadfoot1967}. This leads to blue (428~nm) emissions, which appear as purple or pink against less dark night skies. This process takes place when the upper atmosphere is sunlit. \citet{shiokawa2019} showed that such conditions are met when the solar elevation at the observation site is higher than $-$24$^\circ$ (i.e. less than 24$^\circ$ below the horizon), which corresponds to solar zenith angles comprised between 90$^\circ$ and 114$^\circ$. At $-$24$^\circ$ solar elevation, the shadow of the Earth reaches about 600~km altitude (without accounting for sunlight scattering in the upper atmosphere).

To investigate the origin of the blue and purple/pink/magenta aurora reported by the citizen scientists in the survey, we calculate the solar zenith angle at the time and location of the observations comprising these colours. Figure\,\ref{fig:sza}a shows the solar zenith angle distribution of these observations. Two vertical dashed lines indicate solar zenith angles of 90$^\circ$ and 114$^\circ$, between which the resonant scattering mechanism in the sunlit upper atmosphere (at $\sim$150--400~km) can take place. One can see that the large majority of the reported blue and purple/pink/magenta aurora are associated with solar zenith angles within this range, suggesting that these colours likely find their origin in the resonant scattering mechanism hence requiring low-energy electron precipitation (below 1~keV). Resonant scattering at the top part of the atmosphere allows the vivid pink emission to be seen from a distance. However, this observation alone does not exclude possible higher-energy electron precipitation during the superstorm.The higher-energy electrons would rather results in green emission at lower heights, which may be blocked by the horizon for individual observers viewing from a distance. The fact that the solar zenith angle distribution of green aurora (unfilled green bars) is very similar to that of purple/pink/magenta aurora supports the hypothesis that keV-range electron precipitation was also taking place and visible to the observers closer to the electron precipitation region. This is in agreement with results presented by \citet{Foster2024} who used total electron content measurements along with auroral images in the North-American sector during the Gannon Storm.

While information on the colour of the aurora is not systematically provided in Skywarden observations, many reports and uploaded pictures corresponding to the Gannon Storm contain spectacular displays of blue, purple and pink aurora. Figure\,\ref{fig:sza}b shows the solar zenith angle distribution associated to the Skywarden observations. It is clear that almost all observations were made during conditions suitable for the daylight resonance scattering mechanism to take place. This makes sense, since most of the Skywarden observations are from Finland, where darkness was barely sufficient to see the aurora during the geomagnetic storm.

\begin{figure}[t]
  \includegraphics[width=\textwidth]{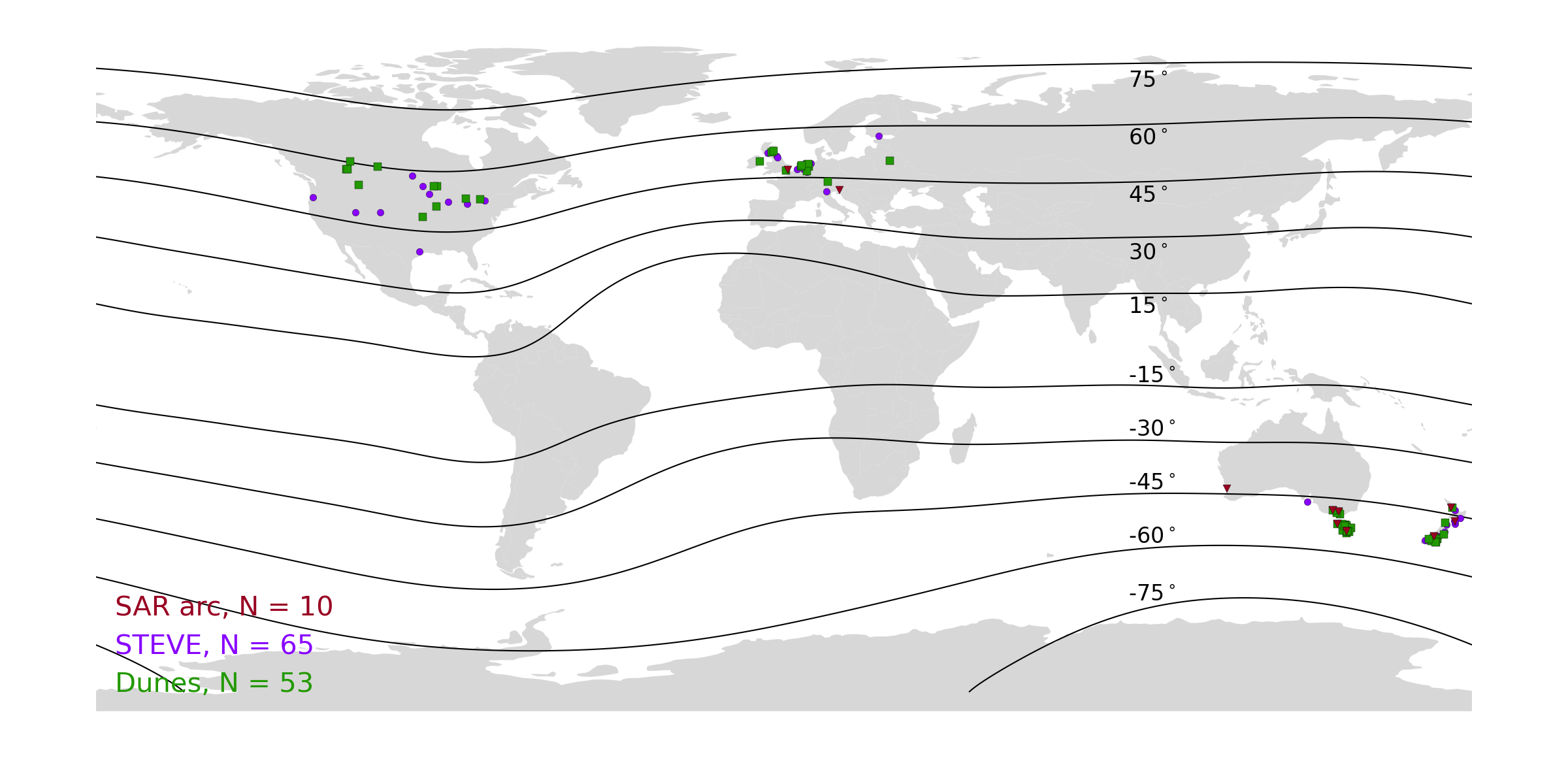}
  \caption{Map of the reports of sightings of SAR arcs (red), STEVE (purple), and the dunes (green).}
  \label{fig:special}
\end{figure}

\subsubsection{Special auroral forms in the reports}

Finally, among the questions asked via the online survey, one was about the observed auroral forms during the event. In the absence of detailed guidance given to citizen scientists regarding the identification and logging of specific auroral forms, analysis of the obtained responses would prove cumbersome and would likely provide little insight. However, we can look at a few optical features which are so peculiar that the risks of mis-identification are low. Figure\,\ref{fig:special} shows the locations from which STEVE (in purple), dunes (in green) and SAR arcs (in red) were spotted during the Gannon Storm. This nomenclature being very specific, it is likely that only citizen scientists already familiar with those phenomena ventured to indicate their presence in the night sky during the event. Overall, 65~reports bear a mention of STEVE, 53~reports indicate the presence of the dunes, and 10~citizen scientists report seeing a SAR arc. One can note that the reports almost exclusively originate from regions where active groups of aurora chasers are present and organised as communities on social media: Canada and the northern United States, the United Kingdom, the Netherlands, Tasmania, and New Zealand. This is consistent with the above statement that only citizen scientists familiar with the concepts of STEVE, dunes and SAR arcs could provide such identifications. While the map given in Fig.\,\ref{fig:special} cannot provide a comprehensive overview of all the regions where these phenomena appeared, nor has it been possible to systematically validate the identifications, it gives a promising indication concerning possible data sources in case one would like to investigate either of these processes in more detail.

\subsection{Disruptions in human-made infrastructure}

\begin{table}[t]
    \centering
    \begin{tabular}{l c}
      \hline
      \textbf{Disruption type} & \textbf{Number of reports} \\
      Geolocalisation problem & 22 \\
      Network/connectivity problem & 11 \\
      Unstable power & 8 \\
      Power cuts & 5 \\
      \hline\\
    \end{tabular}
    \caption{Main types of disruptions during the Gannon Storm as reported by citizen scientists.}
    \label{tab:disruptions}
\end{table}

\begin{figure}[t]
  \includegraphics[width=.5\textwidth]{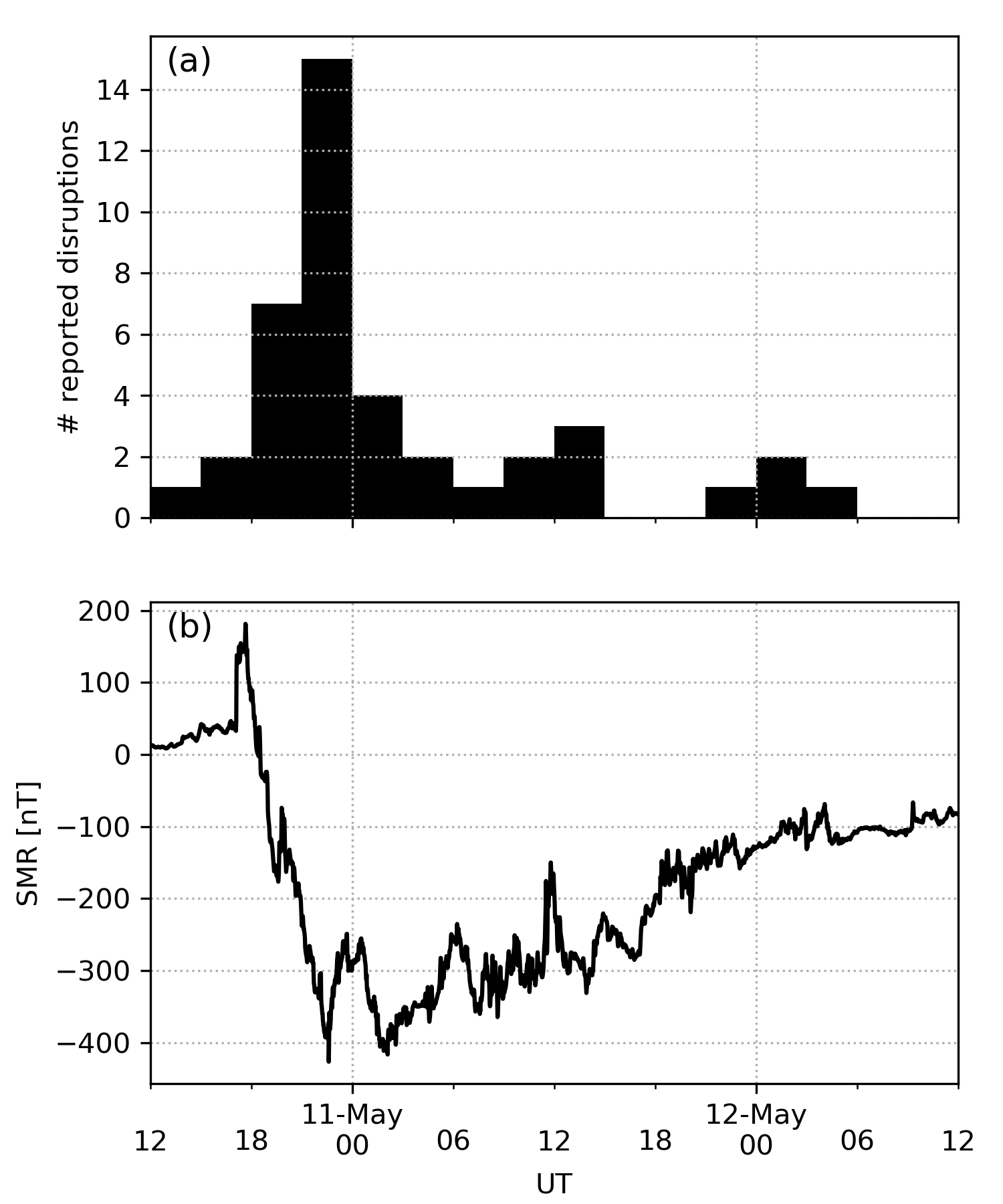}
  \caption{(a)~Distribution of reported disruption times from the survey. (b)~SMR index.}
  \label{fig:disruptions}
\end{figure}

Although the number of disruptions reported via the survey is fairly low (42), it is nonetheless interesting to investigate the types experienced by the citizen scientists during the geomagnetic storm. Table~\ref{tab:disruptions} presents the main types of disruptions reported. Since some citizen scientists reported several disruption types, the sum of the numbers in the right-hand column (46) exceeds the total number of reports (42). More than half of the reports mention issues with geolocalisation services indicating inhomogeneities in the ionospheric E- and F-region electron density producing scintillation of GNSS signals. Such inhomogeneities can be produced by instabilities arising from electron density gradients or turbulence, which can result from particle precipitation \citep[e.g.][]{enengl2024}. About a quarter of the reports indicate problems with network or connectivity; free-text descriptions mention either mobile network or home internet issues. Only a few reports concern disruptions related to electricity (unstable power or power cuts), which may tell about the resilience of our power grids and/or that this storm did not drive extreme current variability in the E region.

It is also instructive to look at the times when disruptions were experienced. Figure\,\ref{fig:disruptions}a shows the distribution of the disruption times. We see that disruptions start after 12~UT on 10~May and the number of reports peaks during the late hours of that day. A second peak of disruption reports takes place after 12~UT on 11~May. In Fig.\,\ref{fig:disruptions}b, we reproduce the SMR time series (same as in Fig.\,\ref{fig:driving}e); we can see that these two peaks correspond well with abrupt decreases in the SMR index, as the geomagnetic storm activity was increasing. However, since the overall number of reports is low, one cannot exclude that the apparent anticorrelation between the number of reports and the SMR index is coincidental. This nonetheless suggests that developing tools to collect disruption reports from citizen scientists during geomagnetic storms could prove important to assess space weather effects on daily life.

\section{Discussion}
\label{sec:discussion}

\subsection{Spontaneously collected data outside of an observation campaign}
We have seen that data collected from citizen scientists can provide insights into space weather events, especially when they are strong enough to affect densely populated mid-latitude regions. In a noteworthy recent publication on the Gannon Storm, \citet{spogli2024} showed examples of aurora observations from the general public in the form of a map shared on Twitter/X along with individual photographs from observers based in Italy and shared via Instagram. They hence highlighted the opportunity given by the superstorm to make communication to the public, which was also one of the motivations of our study. 

The first point to highlight concerning our study is that the collected data were obtained a posteriori as opposed to gathered via a coordinated campaign. This means that the number of reports we obtained could have been significantly greater if there had been a planned campaign of auroral observations, which includes basic training for citizen scientists to know in advance what to look for and how to take pictures of the aurora. Recently, the ARCTICS collaboration released its \emph{Auroral Field Guide and Handbook for Citizen Science} \citep{ARCTICS_handbook} to partly address this challenge. Fortunately, the event was forecast several days in advance, and citizen scientists, photographers, and the general population had been alerted through aurora networks as well as mainstream news outlets. Therefore, this event was observed by many people. As many smartphones now have high-quality cameras, including night modes with longer exposure times, this made it possible for many observers to detect the full range of green, red, blue, and pink emissions which are not typically easy to discern with the naked eye.

\subsection{Opportunity to develop new methods for the analysis of citizen scientist images}
While in this study we did not request aurora pictures from the citizen scientists and mainly presented an analysis of the collected metadata, one should bear in mind that based on the reports logged in via our survey, there exists a large number of auroral images taken from citizen scientists from all around the world for this exceptional event. This opens possibilities for future analysis of those images also in a quantitative manner. 

For instance, new image processing methods can enable the conversion of the red, green and blue (RGB) pixel values from a raw photograph into scientific parameters. Recent simulations of the full synthetic spectra of the aurora performed with the Transsolo code \citep{lummerzheim1994,lilensten2002} allow the estimation of the mean energy of precipitating particles from the ratios between the RGB brightness values. Preliminary calculations using the MATLAB function {\tt spectrumRGB} between 380 and 900~nm (Matlab Central, Juliana Mather 2024, \url{https://www.mathworks.com/matlabcentral/fileexchange/7021-spectral-and-xyz-color-functions}) yield a dominant level in the red channel for low-energy precipitation (in the hundred electronvolt range), comparable RGB levels for medium energies (a few kiloelectronvolts), and a dominant level in the green channel for higher energies (10~keV or more). These colour ratios are considered when looking at the zenith or near-zenith elevations which provides an altitude-integrated view. When looking at low elevations, each pixel from the detector contains signal originating from a limited altitude range, which needs to be accounted for in the analysis. 

In the case of the May 2024 superstorm, the citizen scientist reports suggest that a large part of the pictures show predominantly red emissions, which implies a significant fraction of the precipitating electrons have energies below 1~keV. Applying a systematic analysis of the citizen scientists' images could provide estimates of the precipitating electron mean energies at different times and places during the superstorm, and more complex methods such as tomographic reconstructions \citep{robert2023} could be envisaged and tested.

\subsection{Limitations of the study and lessons learnt}
Among the main limitations of our approach to collect the citizen science data is the difficulty to validate data accuracy. Given that the reports were manually logged by citizen scientists with little guidance and formatting constraints, a non-negligible number of corrections had to be made, especially on observation times, time zone, and geolocalisation (see Sect.~\ref{sec:datacollection}). Besides, some of the collected data proved difficult to analyse and did not yield any insightful results. For instance, the question asking about the location of the aurora in the sky with respect to the observer (see Appendix~\ref{sec:surveyquestions}) contained possible answers (polewards, equatorwards, near zenith, everywhere) that were likely too subjective or too difficult to remember afterwards. Likewise, the question about auroral form identification (quiet arc, moving/active arc, spirals/swirls, nearly vertical beams/pillars, corona, blobs/patches, diffuse aurora, pulsating aurora, dunes, STEVE) proved too technical for many citizen scientists, a significant fraction of whom saw the aurora for the first time in their life during the studied superstorm. 

This underlines the necessity for clearer and more accurate directions for citizen scientists responding to forms such as the one distributed. There should also be an effort to widely distribute accessible documentation for citizen scientists to refer to when trying to identify auroral forms. The ARCTICS Handbook \citep{ARCTICS_handbook}, released a few months after the Gannon Storm occurred, contains guidance and advice on how to take auroral photographs for scientific purposes and how to handle the associated metadata (e.g. time, location, camera settings, etc.). Besides, training sessions for aspiring citizen scientists could be another way to improve the quality and accuracy of the collected data in view of more detailed studies of auroral processes. While the objective of the online survey was not to directly collect the images taken by citizen scientists, it is worth pointing out that dedicated platforms such as Skywarden can readily be used for such purposes, as they are designed to guarantee good practice in terms of data archival, data quality control, content verification, data reusability/access, credit attribution and auroral form classification.

Concerning the reported disruptions on human-made infrastructure, using an online survey or social media posts to collect data comes with even more shortcomings. First, the causality link between a disruption experienced by a user and space weather activity cannot be demonstrated. Although the temporal distribution of the reports shown in Fig.\,\ref{fig:disruptions}a is consistent with the SMR index variations, one cannot exclude that there may be a bias in reporting disruptions that occurred at the same time as strong auroral activity was present, while similar disruptions occurring during geomagnetically quiet times would be left unnoticed (illusory correlation). Second, some of the reported disruptions (e.g. network problems) can have multiple causes, and without being able to investigate each individual case in detail, it is impossible to confirm a possible geomagnetic origin for it. 

Nevertheless, one should not hastily discard the potential of gathering disruption reports from citizen scientists, as this can provide large amounts of data giving insight in space weather effects and evaluate their economic cost. For instance, according to the Solar Influences Data Analysis Centre from the Royal Observatory of Belgium, some GNSS-based tools for aviation were not working properly for several hours during the Gannon Storm, affecting aircraft navigation (\url{https://www.sidc.be/article/extremely-severe-geomagnetic-storm}). Besides, in Canada and the northern United States, there have been reports of farmers being unable to use navigational systems on which their farming equipment rely (e.g. \url{https://www.cbsnews.com/minnesota/news/solar-storms-cause-gps-outage-is-tractors-planting-season/}); a few responses to our survey contained similar statements as free-text comments. For some, this affected the accuracy of the planting of their crops right at the peak of the planting season, which may yield a loss of productivity. Evaluating the economic cost of the superstorm to the farming industry and other sectors could prove useful to devise mitigation strategies in future. This calls for setting up adequate tools to facilitate and systematise reporting, so that the collected data can easily be used in quantitative studies.

\subsection{Going further: extreme storm effects from publicly available data}

\begin{figure}[t]
  \includegraphics[width=.5\textwidth]{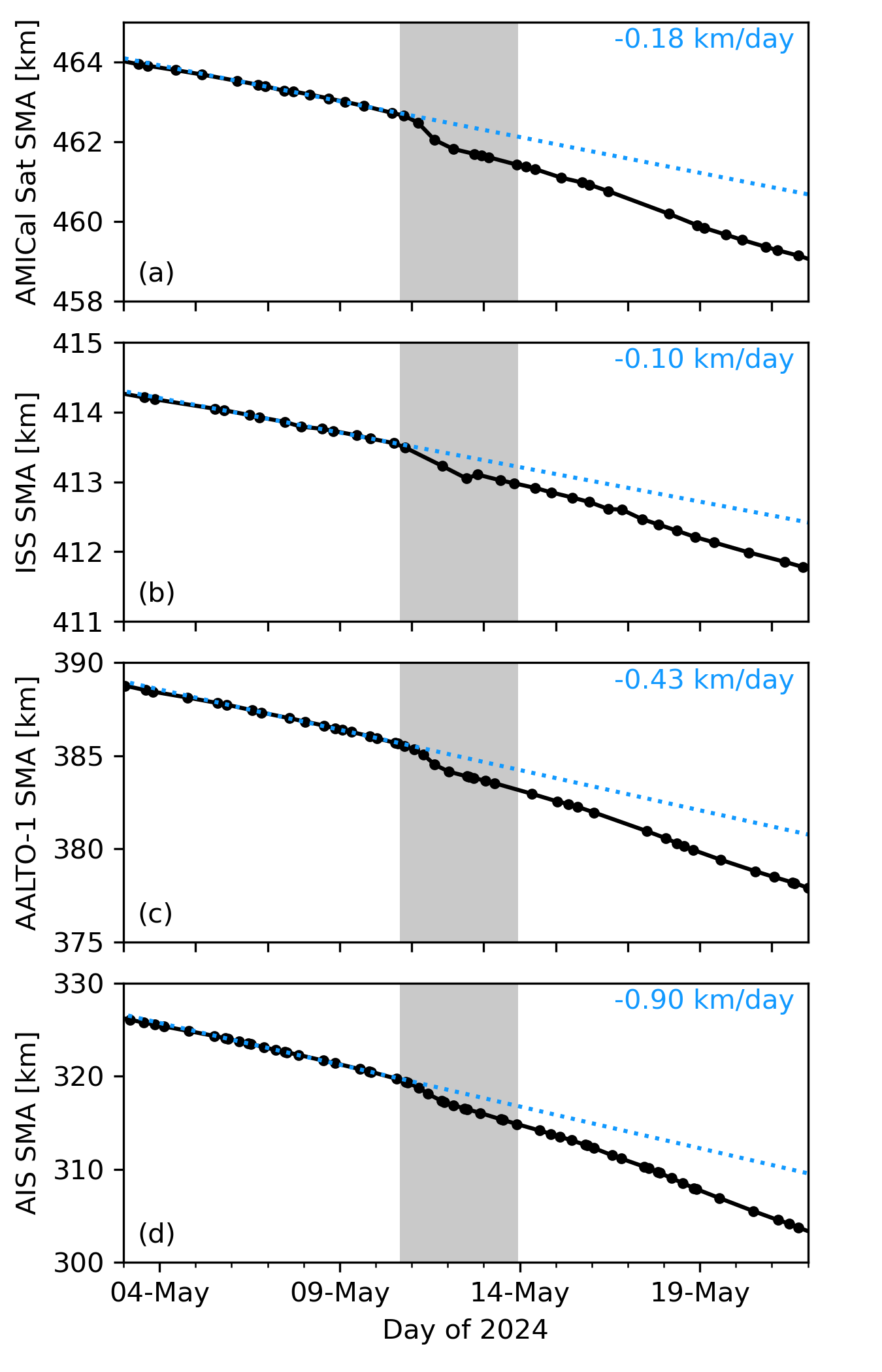}
  \caption{Time evolution of the semi-major axis (SMA) altitude of selected spacecraft in May 2024. The data are shown for (a)~AMICal Sat (NORAD ID \#46287), (b)~the International Space Station (ISS), (c)~AALTO-1 (NORAD ID \#42775), and (d)~AIS \& APRS \& ARIS \& R/B (NORAD ID \#44104). The grey shaded area indicates the superstorm period. The blue dotted lines indicate the extrapolation of the SMA altitude based on the pre-storm altitude loss rate, given in the top-right-hand corner of each panel for the corresponding spacecraft.}
  \label{fig:satellite}
\end{figure}

More generally, not only citizen science but also publicly available data could be more systematically utilised to infer scientific results on space weather effects during extreme geomagnetic events like the Gannon Storm. For instance, orbit data of thousands of satellites can be retrieved via services such as CelesTrak (\url{https://celestrak.org}). When a strong geomagnetic storm takes place, ion--neutral frictional (Joule) heating takes place in the ionospheric E-region within the auroral oval, leading to vertical expansion of the upper atmosphere \citep{palmroth2021}. As a result, low-Earth-orbiting spacecraft experience increased drag, which speeds up their altitude loss. Figure\,\ref{fig:satellite} shows the semi-major axis altitude of the orbits of four selected satellites between 3 and 22~May 2024. The satellites are, by decreasing altitude: AMICal Sat (NORAD ID \#46287), the International Space Station (ISS), AALTO-1 (NORAD ID \#42775), and AIS \& APRS \& ARIS \& R/B (NORAD ID \#44104). The time interval indicated with grey shading, from 16~UT on 10~May until the end of 13~May, corresponds to the superstorm period. 

The orbit data of the four spacecraft exhibit clear signatures during the superstorm, as their semi-major axis altitudes dropped noticeably faster than their quiet-time decay rate shortly after the beginning of the event. To estimate the altitude loss due to the superstorm, we performed a linear fit of the pre-storm data, giving the average quiet-time decay rate of the altitude of each spacecraft in early May 2024 (numbers indicated in blue in the top-right-hand corner of each panel). The linear fits, extrapolated to the superstorm period and until the end of the displayed time interval, are shown with blue dotted lines. We can estimate the altitude loss resulting from the geomagnetic activity during the superstorm by comparing the measured satellite semi-major axis altitude at the end of the storm period with the extrapolated value at that same time. This yields a superstorm-associated altitude loss of 717~m for AMICal Sat, 256~m for the ISS, 1131~m for AALTO-1, and 2091~m for AIS \& APRS \& ARIS \& R/B. A recent study by \citet{parker2024} showed that several thousands of low-Earth-orbiting satellites performed maneuvers simultaneously after the superstorm to compensate for the altitude loss, which is unprecedented.

We can see the effect increases as we consider spacecraft orbiting at lower altitudes, which is a result of the density structure of the upper atmosphere following an exponential decay with altitude. The ISS is an exception to this trend; although its size makes it prone to atmospheric drag, the inclination of its orbit (51.6$^\circ$) confines it at low- and mid-latitudes. This is contrary to the other three satellites which are on polar orbits and hence flew through the entire auroral oval, experiencing enhanced drag along greater portions of their orbits. Nevertheless, it is known that the effects of space weather on satellite drag are not limited to those on polar orbits, as local heating within the auroral zone can be redistributed polewards and equatorwards via upper-atmospheric dynamics \citep{lu2016} and hence affect the mass density at mid-latitudes. Simulations from the empirical US Naval Research Laboratory’s Mass Spectrometer and Incoherent Scatter radar Extended model \citep[NRLMSISE-00,][]{picone2002} indicate significant atmospheric neutral density enhancement at 400~km altitude during the Gannon Storm, also affecting mid-latitudes \citep{parker2024}, supporting this interpretation.

However, in the case of the Gannon Storm, it cannot be excluded that direct heating affected the most poleward portions of the ISS orbit, given that the aurora was seen from locations significantly equatorward from 50$^\circ$ geographic latitude (see Fig.\,\ref{fig:all_reports}). This would be consistent with neutral temperature enhancement observations made by the Global-scale Observations of the Limb and Disk (GOLD) satellite during the event \citep{Evans2024}. In fact, little data on the width of the auroral oval exist for extreme events. While a previous study by \citet{case2016_eqextent} showed that citizen scientist reports of aurora sightings via Aurorasaurus were overall consistent with view line predictions by the OVATION Prime model \citep{newell2014}, this result held for moderate-to-strong geomagnetic activity (Kp\,$\leq$\,7). For extreme events such as the Gannon Storm, models do not predict as broad an oval as what can be inferred from our citizen scientist reports \citep[e.g.][]{sigernes2011,newell2014,blake2021,spogli2024}. This highlights the importance of considering citizen science observations to supplement and improve auroral oval prediction models. Such endeavours would prove crucial to not only better forecast auroral visibility when extreme events are expected, but to also evaluate more finely the extent of the areas which may be affected by space weather events, both on the ground and along satellite orbits.

\conclusions[Conclusions]  
\label{sec:conclusions}

We presented a study of the auroral observations and experienced disruptions during the Gannon Storm based on citizen scientist reports gathered via an online survey. A total of of 696~observers from around the world submitted responses which enabled getting insights in the effects of the superstorm that could not be obtained with the existing scientific infrastructure alone.

Lessons learnt from this study can be summarised as follows:
\begin{enumerate}
    \item Communities of aurora chasers or enthusiasts following dedicated social media pages are likely to actively get involved in citizen science, even when having had no prior experience with it. Although few people consider themselves avid citizen scientists, the general public will respond to a request from a trusted leader or group that they follow, especially when there is an incentive such as getting their observation reported on a map or analysed in a study. This underlines the importance of collaborating with community managers or ambassadors who can act as intermediaries between academics and the general public.
    \item The collection and pre-processing of citizen science data poses challenges, such as outlining adequate guidance and background information for citizen scientists to be able to answer any questions asked of them and constraining the format in which data are logged. This highlights the need for openly available documentation for citizen scientists, as well as the crucial role of data collection platforms managed according to professional data collection practices, such as Skywarden.
    \item The analysis of the collected aurora sightings during the Gannon Storm showed that the aurora was visible down to beyond 40$^\circ$ geomagnetic latitude with a few reports as low as below 30$^\circ$. Auroral oval boundary models are currently not capable of accounting for observations at such low geomagnetic latitudes during extreme events.
    \item The colours of aurora reported by the citizen scientists included unusually bright reds, pinks and blues, even visible to the naked eye. This suggests that large fluxes of low-energy (<\,1~keV) electron precipitation were dominant during those phases of the event. The pink hues can be explained by high-altitude blue emissions in the sunlit upper atmosphere.
    \item Only a few reports on disruptions were made, yet their temporal distribution matched well with the geomagnetic activity level as measured by the SMR index. The main types of experienced disruptions were issues with geolocalisation systems, network, and power. Dedicated tools to collect reports on human-made infrastructure disruptions during geomagnetic events could prove valuable in future cost assessment due to space weather disturbances.
\end{enumerate}

The Gannon Storm is exceptional in terms of aurora reports, especially those from low latitudes beyond predictions from current auroral oval models. Besides, perturbations on GNSS signals were reported from various latitudes, and low-Earth-orbiting satellites lost up to 2~km in their semi-major axis altitude due to the geomagnetic activity. While thankfully no major power outage or spacecraft losses were reported during this superstorm, presumably because of the predominantly low-energy particle precipitation, major space weather events with different driving conditions could have much more devastating consequences. Therefore, it is crucial to make use of all available data to ensure that effects and processes taking place at latitudes not covered by scientific instruments are measured and characterised. Thanks to the enthusiasm of citizen scientists all over the world, such data can be collected during future events. Improving the citizen science data collection and management methods is paramount to enable more quantitative studies and monitoring of the Earth environment.



\vspace*{0.5cm}
\small{\noindent\emph{Data availability.} Solar wind data were downloaded from OMNIWeb at \url{https://omniweb.gsfc.nasa.gov}, SuperMAG magnetic index data are available at \url{https://supermag.jhuapl.edu}, and spacecraft orbit altitudes can be downloaded from CelesTrack at \url{https://celestrak.org}. The VIIRS DNB image shown as the background of Fig.~\ref{fig:oval} was obtained from \url{https://cimss.ssec.wisc.edu/satellite-blog/archives/59112}. The Skywarden data were retrieved from \url{https://www.taivaanvahti.fi/} via the API (direct download link for the superstorm time period: \url{https://www.taivaanvahti.fi/app/api/search.php?format=csv&start=2024-05-10&end=2024-05-13&cat=2&columns=all&language=en}). The data generated via the online survey have been deposited -- after anonymisation as described in Sect.~\ref{sec:datacollection} -- on Zenodo \citep{survey_data}, along with their documentation.} 




\clearpage
\appendix
\section{Survey Questions} \label{sec:surveyquestions}
The following is the list of questions citizen scientists answered in the survey:
\begin{itemize}
    \item Name
    \item Do you want your name to be acknowledged in the report?\
    \item Email address (if you want to be updated on the report)
    \item Location (closest town/city, state/region/province \& country)
    \item Geographic (GPS) coordinates, if you know them (e.g., 38.898, -77.036)
    \item Did you experience any disruption (e.g. geolocalisation issues, power cuts, flickering lights...) during the storm?
    \begin{itemize}
        \item[] \textit{If answered yes to the above question:}
        \item Date when you experienced the disruption
        \item Approximate time when you experienced the disruption
        \item Time zone in which the above time is given
        \item If you experienced disruptions several times during the storm period, you can list them below.
        \item What kind(s) of disruption(s) did you experience?
    \end{itemize}
    \item Did you see the aurora during this event?
    \begin{itemize}
        \item[] \textit{If answered yes to the above question:}
        \item Date when you saw/photographed the aurora
        \item Approximate time at which you saw/photographed the aurora (if you stayed for an extended period, indicate the starting time)
        \item Time zone in which the above time is given
        \item If you saw/photographed the aurora several times during the storm period, you can list them below.
        \item Location where you saw the aurora (city, state/region/province \& country, or GPS coordinates)
        \item What type(s) of aurora did you see?
        \item What colors of aurora did you see (naked eye)?
        \item What colors of aurora did you capture (in pictures)?
        \item Location of the aurora relative to you
    \end{itemize}
\end{itemize}

\section{List of the data providers (online survey)}    
\label{sec:csnames}
The data presented in this study were collected via an online form between 18~May and 24~June 2024. A total of 696~citizen scientists filled in a report on their observations during the superstorm, among whom 195 wished to remain anonymous. The 501~contributors who consented to being acknowledged by name are: Ada Verbree-Guijt, Adam Meyer, Adeline Sarah Jebaraj, Adriana Kasel, Aishath Zahwa, Aishwerya Kapoor, Akshatha Gopinath, Alan Dyer, Alasdair Taylor, Alessandro, Aletta Elders, Alex Daykin, Alexander, Alexander, Alexis Clift, Alfredo Juarez, Alison Thomass, Allison Mills, Amanda, Amanda, Amanda Hawn, Amrit Grewal, Amy Hester, Amy Willis, Anais Giblot-Ducray, Andrea McArthur, Andreas Milanese, Andrew Cadie, Andrew Vis, André Magalhães, Angela Gnewikow, Angela Nicol, Angela Robinson, Anke Verhaegen, Ann E. Johnson, Ann Towers, Anna Gottlieb, Anne Rodriguez, Anne Verzijl, Anne-Sophie Bescond, Anouk Vorselman, Anoula Voerman, Arjan, Arjan Kievits, Arjan de Jong, Arlene Oetomo, Arnaldo Lopez, Ashley Elsworth, Astrid Broere, Audrey Todd, Austin E., Ayla Embil, B. Wassenaar, Barbara koster, Barend (Barry) Becker, Bas Zonneveld, Benjamin Barakat, Berna van Tol, Beth Mason, Bo Krause, Boris Baloh, Brandon Walsh, Brighton Seeley, Bryn Jones, Candace Montgomery, Carlo W., Carmen Varela, Carola van Hof, Carole Sneed, Caroline Meijer, Caroline Whittaker, Carrie Bastyr, Centro de Instrumentación Científica de UNACH, Charles Le Béhot, Charlotte Bridgett, Charmayne, Chloe Jellis, Chrissey, Christa Creech, Christian Harris, Christina, Christopher Jones, Christos Doudoulakis, Chuck Benz, Claire Doran, Claudia Koekoek, Claudia Meulenbelt, Clément, Cole Gifford, Colette Dupont, Colin Legg, Constance Lewis, Corine, Corinna Zeller, Corné Ouwehand, Cory Goings, Croydon Hall, Daan van Laar, Dai Jianfeng, Daisy, Dakota Snider, Dale Turner, Dale Weigt, Dalton Mojica, Damijan Prosenak, Dan Nigro, Daniel Fernandes Gama, Danielle ODea, David Hunter, David Lusby, David Rius Serra, David Roberts, David Smith, Davide Maligno, Dawn Burbidge, Deb Angel, Deb Dennison, Deb Wiensch, Debbie, Debra Pierrehumbert, Diana Marell, Diane Chandler, Diane Thompson, Dmitri Kamenetsky, Donna Lach, Dorien ten Hulscher, Doug Cottrell, Eero Karvinen, Elan Azriel, Elizabeth Miller, Elizabeth Palmer, Eman Dannawey, Embla Wihk, Emily Donley, Emma Bruus, Enrique Carrasco, Entoni Novosel, Enzo Carlos, Eren Matthews, Eric, Eric B., Erin Dijkstra, Ernstjan Penninkhof, Erum Tanvir, Esther, Eva van Gent, Ewan Kane, Ezgi Gülay, Farin Drewes, Federico Butac, Ferdie, François Le Béhot, Freja van der Niet, Gail Willenbring, Gaynor, Georgina Malisauskas, Gerrie, Gerry Buckel, Gert van Eck, Gertjan van Norel, Godfried Nijs, Graeme Whipps, Graham Russell, Grant Birley, Gregoire Deprez, Guillaume Laget, Hajare Ait Ouaba, Hans-Maarten, Harry Selwood, Hayley Young, Helen Ruiter, Hetty, Holly Matter, Howard Cheng, Ian Cooper, Ian Wiseman, Ineke, Ineke van Poelgeest, Ines Jonkers, Iona Baggerman, Ipek G. Kulahci, Isabel Kessler, Ivica Skokic, Jack Delves, Jacqueline Horkings, Jake Winters, Jamie McBean, Jan Twomet, Jane Herbert, Janeth Vargas, Jannie Bouwman, Jason Kurth, Jasper den Hollander, Javier Caldera Fernández de los Muros, Jaxon Hoffmann, Jayme, Jen Browning, Jen Makin, Jen McDonald, Jendy Jalving, Jennifer, Jennifer Babin, Jennifer Bailey, Jennifer Docherty, Jennifer Falkofske, Jennifer Flor, Jennifer Giesbrecht, Jennifer Hendricks, Jennifer McIntosh, Jenny Atkinson, Jenny Clark, Jeremy Kuzub, Jeremy Mion, Jeremy Perez, Jerry Eklund, Jessica, Jessica Dorsey, Jessica Keller, Jessica Leanne, Jessica Pruter, Jessica Reurich, Jessica Taylor, Jill Silverberg, Joanna Fox, Joanne Kampinga, Jodan, Jodi Baker, Joe Cali, Joelka van Daal, John Oxley, Joke Foppen, Jonas Suni, Jose Luis Hormaechea, Jose Reynaga, Josefine Friedemann, Julia Sumerling, Julian Soldat, Julie Hayes, Julie Lawson, Jure Atanackov, Jørgen van Meijbeek, Kaetlyn, Kaitlyn Krus, Kaitlynn Williams, Kali Salmas, Karin Biermans, Kat Smith, Kate Cameron, Kate Draeger, Katharina Amon, Katherine, Kathy Janssen-Gubbels, Kathy Olson, Katie Giles, Katie Moloney, Katie Raymer-Woods, Kaveendra Daluwathumulla, Kayla, Kees Zwaan, Kellie Louise, Kelsey Harms, Ken, Kendall K., Kerri Johnson, Kevin, Kevin Kelleher, Kevin Palmer, Kezia Kurian, Khoi Nguyen, Kieran, Kim Hesse, Kimberly Shorkey, Kirsten Steele, Klaas Jobse, Kori Gill-Davidson, Kristina Young, Kylie Gee, Kyra Wing, Laura Hansen, Lauren Guenon, Laurie Crofoot, Lee K., Les Ladbrook, Leslie Mouncey, Liane Henry, Libby Gabriel, Lilian Kars, Lily Neyland, Linda, Linda S., Lindey, Lindsay Eastman, Lisa K. Hyatt, Lisa Starr, Liz Halliday, Liz Stuart, Loes Aartsen, Logan, Lone, Lotte Enting, Luc Jeanjean, Luke Rasmussen, Luke Verschoor, Lynn Sosnoskie, MJ van Hengel, Madalyn Draper, Malcolm Park, Mandy Erades, Manja, Manon Véber, Marcia Boomhouwer, Marco, Margaret Fitzgerald, Margaret Sonnemann, Margot Koorenhof, Marianne, Marianne Coppens, Marianne van den Berg, Marjan Spijkers, Marjon Wensink, Mark Egan, Mark N., Martha Loeppky, Martin van Marion, Martina, Martine, Martyn Lloyd, Marybeth Kiczenski, Maryla Machlick, Matthew Gaines, Matthew Pfab, Maxy, Meg, Megan Marsh, Melanie Clarke, Melanie Fama, Melissa, Melissa Kaelin, Merlijn, Mia Noyens, Michael Harmon, Michael Tomeh, Michelle, Michelle Bandy, Michelle Maynard, Michelle Stephenson, Mike Smith, Mirjam Jansen, Mirjam Leyte, Mirriam, Missy McCormick, Monique van Oijen, Monique, Montse, Morgan merrell, Nadia Tildesley, Nadja, Nanette Smith, Narelle, Natalie Pace, Natalie Sinkr, Natasha Peiskar, Nate Avish, Nepal Nelson Palma, Nicholas, Nicola Jackson, Nicolas Achmadi, Nicole Anderson, Nicole, Nicole P., Nicole Walton, Nicolás Concha Vargas, Nienke Huijsing, Nik Zimmermann, Nikita Loreggian, Nikki Boys, Nikki Dayton-Gelati, Niko, Niké Spits, Nory, Obbe, Oliver Saunders Wilder, Padmini Selvaganesan, Paige West, Pam \& Andy Barnes, Pandora Biskner, Parmeet Singh, Pat Russell, Patricia, Patrick Johanneson, Paul Brooks, Paul West, Paul Williams, Paula Mair, Paulina, Pearl Wong, Peter Dohnt, Peter R. Grounds, Petra Bloemsma, Piet Berger, Pip Reisch, Pippa Mitchell, Poppy Franklin, Rachael Ramirez, Rachel Baxter, Randy Milanovic, Raul Saavedra, Raylene Garwood, Rebekah DeVinney, Remus, René Wolf, Ricardo Panuzzio, Ricardo Velasco, Richard Pyne, Ricki-Lee Teague, Rita Baker, Robbi James, Robert Fear, Robert-Jan, Robin Moon, Rolinde Hatzmann, Rona Hatzmann-Jorritsma, Roope Hakanpää, Rosie Cooper, Rosie Johnson, Ross Johnston, Rudy Siggs, Ryan Latterell, Ryan Voutilainen, Sabrina Spanjaart, Sandi, Sandra Alder, Sandra Stemmerik, Sankalp Merchant, Santosh Ramaswamy, Sarah H. Taft, Seamus FR, Serpollier Cléa, Shannon Landles, Shardae Ros, Sharon, Shaula Corr, Shawn Rosinski, Shawn Saito, Shellie Evans, Sherri Yezbick-Taylor, Shikhar Gupta, Shirley Jones, Sia Nikolaou, Silver Moon, Simon Evans, Sjon de Mol, Sonja Yearsley, Spencer Plovie, Stefan Ayto, Steve Brown, Steve DuBois, Sue Noble-Adams, Sunny Yang, Supriya, Susan Padfield, Susan Snow, Suze, Sylvia de Raat, Tammi Turner, Tammy Vallee, Tanya Melnik, Thomas Grandin, Théo Vischel, Tiffany Lamb, Titouan Joulain, Tom Warner, Tracey Parks, Traci Krzyzanowski, Tracie Blackburn, Travis Vander Laan, Tristan Pokornyi, Ursula Ram, Vaclav, Vanessa Wenners, Victoria Coleman, Vincent Ledvina, Vincent Morand, Virginia, Ward Van Herck, Wendy, Wendy Bakker-Coppens, Wendy Forsyth, Wendy Heugens, Wesley Souza, William McQuillan, Xavier Meneboode, Yolanda de Groot, Ytsje Kobus, Yvette Phillips, Yvo Ambags, Yvonne Heijboer, Zade Johnston, Zeel Parmar, Zia Self, Zoe Alexander.

\noappendix       




\appendixfigures  

\appendixtables   


\vspace*{0.5cm}
\small{\noindent\emph{Author contributions.} MG and BGL set up the online survey. MS, DL, VEL, RDO, EB, EK, MB, CM, BGL, and MG contributed to sharing the survey as broadly as possible. EB assisted in the collection of the Skywarden data; MB assisted in the collection of CelesTrak data. VEL prepared Fig.\,\ref{fig:oval}. MG pre-processed and analysed the collected data and wrote most of the manuscript text, with inputs from EB, VEL, NP, MB, CM, RDO, and YN. All authors have read and provided feedback on the manuscript.} 

\vspace*{0.5cm}
\small{\noindent\emph{Competing interests.} The authors declare that they do not have any conflict of interests.} 


\vspace*{0.5cm}
\small{\noindent\emph{Ethical statement.} The citizen science data collection was made via an online survey, which has been reviewed by the Data Management Services of the University of Helsinki. It complies with EU's GDPR, and the privacy notice associated to the data management is linked in the data description file. The dataset published as a result of the analysis of the collected data has been anonymised and contains no personal data.}

\vspace*{0.5cm}
\small{\noindent\emph{Acknowledgements.} 
The authors gratefully thank the 696~citizen scientists who answered the online survey between 18~May and 24~June 2024, making this study possible.  The names of those who did not wish to remain anonymous are listed in Appendix~\ref{sec:csnames}. We also wish to thank Colin Legg, Les Ladbrook, and Margaret Sonnemann for sharing the survey link to their communities and giving feedback on the manuscript. In addition, the authors wish to thank the 148~named and 22~anonymous citizen scientists who contributed additional data to the analysis through the Skywarden observation system.
This research was supported by the International Space Science Institute (ISSI) in Bern, Switzerland, through ISSI Working Group project ARCTICS. We gratefully acknowledge the SuperMAG collaborators (http://supermag.jhuapl.edu/info/?page=acknowledgement). We acknowledge use of NASA/GSFC's Space Physics Data Facility's OMNIWeb service, and OMNI data. The work of MG is funded by the Research Council of Finland (grants 338629-AERGELC'H and 360433-ANAON). Financial support is provided to KH by the Research Council of Norway under contract 343302. The work of YN was supported by NASA grants 80NSSC21K1321, 80NSSC23K0410, and 80NSSC23M0193, NSF grant AGS-1907698, and AFOSR grants FA9550-23-1-0614. CM is supported by NSF Aeronomy grant \#AGS-2152365. EK acknowledges the Magnus Ehrnrooth Foundation for a travel grant to attend the ISSI meeting on 3--7~June 2024 in Bern, and DL thanks the University of Calgary for travel support to attend that meeting.
}


\begin{thebibliography}{71}
\providecommand{\natexlab}[1]{#1}
\providecommand{\url}[1]{\texttt{#1}}
\providecommand{\urlprefix}{}
\expandafter\ifx\csname urlstyle\endcsname\relax
  \providecommand{\doi}[1]{https://doi.org/\discretionary{}{}{}#1}\else
  \providecommand{\doi}{https://doi.org/\discretionary{}{}{}\begingroup
  \urlstyle{rm}\Url}\fi

\bibitem[{Akasofu(2005)}]{akasofu2005}
Akasofu, S.-I.: Long-standing unsolved problems in solar physics and
  magnetospheric physics, in: Multiscale Coupling of Sun-Earth Processes,
  edited by Lui, A.~T.~Y., Kamide, Y., and Consolini, G., pp. 71--82, Elsevier
  Science B.V., Amsterdam, ISBN 978-0-444-51881-1,
  \doi{https://doi.org/10.1016/B978-044451881-1/50006-X}, 2005.

\bibitem[{{Alfonsi} et~al.(2022){Alfonsi}, {Bergeot}, {Cilliers}, {De
  Franceschi}, {Baddeley}, {Correia}, {Di Mauro}, {Enell}, {Engebretson},
  {Ghoddousi-Fard}, {H{\"a}ggstr{\"o}m}, {Ham}, {Heygster}, {Jee}, {Kero},
  {Kosch}, {Kwon}, {Lee}, {Lotz}, {Macotela}, {Marcucci}, {Miloch}, {Morton},
  {Naoi}, {Negusini}, {Partamies}, {Petkov}, {Pottiaux}, {Prikryl},
  {Shreedevi}, {Slapak}, {Spogli}, {Stephenson}, {Triana-G{\'o}mez},
  {Troshichev}, {Van Malderen}, {Weygand}, and {Zou}}]{alfonsi2022}
{Alfonsi}, L., {Bergeot}, N., {Cilliers}, P.~J., {De Franceschi}, G.,
  {Baddeley}, L., {Correia}, E., {Di Mauro}, D., {Enell}, C.-F., {Engebretson},
  M., {Ghoddousi-Fard}, R., {H{\"a}ggstr{\"o}m}, I., {Ham}, Y.-b., {Heygster},
  G., {Jee}, G., {Kero}, A., {Kosch}, M., {Kwon}, H.-J., {Lee}, C., {Lotz}, S.,
  {Macotela}, L., {Marcucci}, M.~F., {Miloch}, W.~J., {Morton}, Y.~J., {Naoi},
  T., {Negusini}, M., {Partamies}, N., {Petkov}, B.~H., {Pottiaux}, E.,
  {Prikryl}, P., {Shreedevi}, P.~R., {Slapak}, R., {Spogli}, L., {Stephenson},
  J., {Triana-G{\'o}mez}, A.~M., {Troshichev}, O.~A., {Van Malderen}, R.,
  {Weygand}, J.~M., and {Zou}, S.: {Review of Environmental Monitoring by Means
  of Radio Waves in the Polar Regions: From Atmosphere to Geospace}, Surveys in
  Geophysics, 43, 1609--1698, \doi{10.1007/s10712-022-09734-z}, 2022.

\bibitem[{{Archer} et~al.(2019){Archer}, {St. -Maurice}, {Gallardo-Lacourt},
  {Perry}, {Cully}, {Donovan}, {Gillies}, {Downie}, {Smith}, and
  {Eurich}}]{archer2019}
{Archer}, W.~E., {St. -Maurice}, J.~P., {Gallardo-Lacourt}, B., {Perry}, G.~W.,
  {Cully}, C.~M., {Donovan}, E., {Gillies}, D.~M., {Downie}, R., {Smith}, J.,
  and {Eurich}, D.: {The Vertical Distribution of the Optical Emissions of a
  Steve and Picket Fence Event}, Geophysical Research Letters, 46,
  10,719--10,725, \doi{10.1029/2019GL084473}, 2019.

\bibitem[{{Baker} et~al.(2013){Baker}, {Li}, {Pulkkinen}, {Ngwira}, {Mays},
  {Galvin}, and {Simunac}}]{Baker2013}
{Baker}, D.~N., {Li}, X., {Pulkkinen}, A., {Ngwira}, C.~M., {Mays}, M.~L.,
  {Galvin}, A.~B., and {Simunac}, K.~D.~C.: {A major solar eruptive event in
  July 2012: Defining extreme space weather scenarios}, Space Weather, 11,
  585--591, \doi{10.1002/swe.20097}, 2013.

\bibitem[{{Bates}(1949)}]{bates1949}
{Bates}, D.~R.: {The Intensity Distribution in the Nitrogen Band Systems
  Emitted from the Earth's Upper Atmosphere}, Proceedings of the Royal Society
  of London Series A, 196, 217--250, \doi{10.1098/rspa.1949.0025}, 1949.

\bibitem[{Bhatt et~al.(2023)Bhatt, Harding, Makela, Navarro, Lamarche,
  Valentic, Kendall, and Venkatraman}]{bhatt2023}
Bhatt, A.~N., Harding, B.~J., Makela, J.~J., Navarro, L., Lamarche, L.~J.,
  Valentic, T., Kendall, E.~A., and Venkatraman, P.: MANGO: An Optical Network
  to Study the Dynamics of the Earth's Upper Atmosphere, Journal of Geophysical
  Research: Space Physics, 128, e2023JA031\,589,
  \doi{https://doi.org/10.1029/2023JA031589}, 2023.

\bibitem[{{Blake} et~al.(2021){Blake}, {Pulkkinen}, {Schuck}, {Glocer}, and
  {T{\'o}th}}]{blake2021}
{Blake}, S.~P., {Pulkkinen}, A., {Schuck}, P.~W., {Glocer}, A., and {T{\'o}th},
  G.: {Estimating Maximum Extent of Auroral Equatorward Boundary Using
  Historical and Simulated Surface Magnetic Field Data}, Journal of Geophysical
  Research (Space Physics), 126, e28\,284, \doi{10.1029/2020JA028284}, 2021.

\bibitem[{{Broadfoot}(1967)}]{broadfoot1967}
{Broadfoot}, A.~L.: {Resonance scattering by N$_{2}^{+}$}, Planetary and Space
  Science, 15, 1801--1815, \doi{10.1016/0032-0633(67)90017-7}, 1967.

\bibitem[{{Bruus}(2024)}]{bruus2024}
{Bruus}, E.: Taivaanvahti/Himlakollen/Skywatcher's search interface,
  \urlprefix\url{https://www.taivaanvahti.fi/app/docs/interface/output_interface_en.html},
  (last accessed: 9 July 2024), 2024.

\bibitem[{Burrell et~al.(2020)Burrell, van~der Meeren, and Laundal}]{aacgmv2}
Burrell, A., van~der Meeren, C., and Laundal, K.~M.: {aburrell/aacgmv2: Version
  2.6.0 [Software]}, Zenodo, \doi{10.5281/zenodo.3598705}, 2020.

\bibitem[{{Case} et~al.(2016{\natexlab{a}}){Case}, {Kingman}, and
  {MacDonald}}]{case2016_alerts}
{Case}, N.~A., {Kingman}, D., and {MacDonald}, E.~A.: {A real-time hybrid
  aurora alert system: Combining citizen science reports with an auroral oval
  model}, Earth and Space Science, 3, 257--265, \doi{10.1002/2016EA000167},
  2016{\natexlab{a}}.

\bibitem[{{Case} et~al.(2016{\natexlab{b}}){Case}, {MacDonald}, and
  {Viereck}}]{case2016_eqextent}
{Case}, N.~A., {MacDonald}, E.~A., and {Viereck}, R.: {Using citizen science
  reports to define the equatorial extent of auroral visibility}, Space
  Weather, 14, 198--209, \doi{10.1002/2015SW001320}, 2016{\natexlab{b}}.

\bibitem[{Donovan et~al.(2006)Donovan, Mende, Jackel, Frey, Syrjäsuo,
  Voronkov, Trondsen, Peticolas, Angelopoulos, Harris, Greffen, and
  Connors}]{Donovan2006}
Donovan, E., Mende, S., Jackel, B., Frey, H., Syrjäsuo, M., Voronkov, I.,
  Trondsen, T., Peticolas, L., Angelopoulos, V., Harris, S., Greffen, M., and
  Connors, M.: The THEMIS all-sky imaging array—system design and initial
  results from the prototype imager, Journal of Atmospheric and
  Solar-Terrestrial Physics, 68, 1472--1487,
  \doi{https://doi.org/10.1016/j.jastp.2005.03.027}, 2006.

\bibitem[{{Eastwood} et~al.(2018){Eastwood}, {Hapgood}, {Biffis}, {Benedetti},
  {Bisi}, {Green}, {Bentley}, and {Burnett}}]{eastwood2018}
{Eastwood}, J.~P., {Hapgood}, M.~A., {Biffis}, E., {Benedetti}, D., {Bisi},
  M.~M., {Green}, L., {Bentley}, R.~D., and {Burnett}, C.: {Quantifying the
  Economic Value of Space Weather Forecasting for Power Grids: An Exploratory
  Study}, Space Weather, 16, 2052--2067, \doi{10.1029/2018SW002003}, 2018.

\bibitem[{{Enengl} et~al.(2024){Enengl}, {Spogli}, {Kotova}, {Jin}, {Oksavik},
  {Partamies}, and {Miloch}}]{enengl2024}
{Enengl}, F., {Spogli}, L., {Kotova}, D., {Jin}, Y., {Oksavik}, K.,
  {Partamies}, N., and {Miloch}, W.~J.: {Investigation of Ionospheric
  Small-Scale Plasma Structures Associated With Particle Precipitation}, Space
  Weather, 22, e2023SW003\,605, \doi{10.1029/2023SW003605}, 2024.

\bibitem[{Evans et~al.(2024)Evans, Correira, Lumpe, Eastes, Gan, Laskar, Aryal,
  Wang, Burns, Beland, Cai, Codrescu, England, Greer, Krywonos, McClintock,
  Plummer, and Veibell}]{Evans2024}
Evans, J.~S., Correira, J., Lumpe, J.~D., Eastes, R.~W., Gan, Q., Laskar,
  F.~I., Aryal, S., Wang, W., Burns, A.~G., Beland, S., Cai, X., Codrescu, M.,
  England, S., Greer, K., Krywonos, A., McClintock, W.~E., Plummer, T., and
  Veibell, V.: GOLD Observations of the Thermospheric Response to the 10–12
  May 2024 Gannon Superstorm, Geophysical Research Letters, 51,
  e2024GL110\,506, \doi{https://doi.org/10.1029/2024GL110506}, 2024.

\bibitem[{{Fang} et~al.(2010){Fang}, {Randall}, {Lummerzheim}, {Wang}, {Lu},
  {Solomon}, and {Frahm}}]{fang2010}
{Fang}, X., {Randall}, C.~E., {Lummerzheim}, D., {Wang}, W., {Lu}, G.,
  {Solomon}, S.~C., and {Frahm}, R.~A.: {Parameterization of monoenergetic
  electron impact ionization}, Geophysical Research Letters, 37, L22\,106,
  \doi{10.1029/2010GL045406}, 2010.

\bibitem[{Foster et~al.(2024)Foster, Erickson, Nishimura, Zhang, Bush, Coster,
  Meade, and Franco-Diaz}]{Foster2024}
Foster, J.~C., Erickson, P.~J., Nishimura, Y., Zhang, S.~R., Bush, D.~C.,
  Coster, A.~J., Meade, P.~E., and Franco-Diaz, E.: Imaging the May 2024
  Extreme Aurora With Ionospheric Total Electron Content, Geophysical Research
  Letters, 51, e2024GL111\,981, \doi{https://doi.org/10.1029/2024GL111981},
  2024.

\bibitem[{{Gjerloev}(2012)}]{SuperMAG_technique}
{Gjerloev}, J.~W.: {The SuperMAG data processing technique}, Journal of
  Geophysical Research (Space Physics), 117, A09\,213,
  \doi{10.1029/2012JA017683}, 2012.

\bibitem[{{Grandin} and {ARCTICS collaboration}(2024)}]{survey_data}
{Grandin}, M. and {ARCTICS collaboration}: {Citizen Science Reports on Aurora
  Sighting and Technological Disruptions during the 10 May 2024 Geomagnetic
  Storm – ARCTICS Survey [Data set]}, \doi{10.5281/zenodo.12732615}, 2024.

\bibitem[{{Grandin} et~al.(2021){Grandin}, {Palmroth}, {Whipps}, {Kalliokoski},
  {Ferrier}, {Paxton}, {Mlynczak}, {Hilska}, {Holmseth}, {Vinorum}, and
  {Whenman}}]{grandin2021}
{Grandin}, M., {Palmroth}, M., {Whipps}, G., {Kalliokoski}, M., {Ferrier}, M.,
  {Paxton}, L.~J., {Mlynczak}, M.~G., {Hilska}, J., {Holmseth}, K., {Vinorum},
  K., and {Whenman}, B.: {Large-Scale Dune Aurora Event Investigation Combining
  Citizen Scientists' Photographs and Spacecraft Observations}, AGU Advances,
  2, e00\,338, \doi{10.1029/2020AV000338}, 2021.

\bibitem[{Green and Boardsen(2006)}]{green2006}
Green, J.~L. and Boardsen, S.: Duration and extent of the great auroral storm
  of 1859, Advances in Space Research, 38, 130--135,
  \doi{https://doi.org/10.1016/j.asr.2005.08.054}, the Great Historical
  Geomagnetic Storm of 1859: A Modern Look, 2006.

\bibitem[{{Greshko}(2024)}]{greshko2024}
{Greshko}, M.: {Extreme solar storm generated aurorae—and ‘surprise’},
  Science Insider, 13 May 2024, \doi{10.1126/science.zoakd5z}, 2024.

\bibitem[{Gritsevich et~al.(2014)Gritsevich, Lyytinen, Moilanen, Kohout,
  Dmitriev, Lupovka, Midtskogen, Kruglikov, Ischenko, Yakovlev, Grokhovsky,
  Haloda, Halodova, Peltoniemi, Aikkila, Taavitsainen, Lauanne, Pekkola, Kokko,
  and Larionov}]{gritsevich2014}
Gritsevich, M., Lyytinen, E., Moilanen, J., Kohout, T., Dmitriev, V., Lupovka,
  V., Midtskogen, S., Kruglikov, N., Ischenko, A., Yakovlev, G., Grokhovsky,
  V., Haloda, J., Halodova, P., Peltoniemi, J., Aikkila, A., Taavitsainen, A.,
  Lauanne, J., Pekkola, M., Kokko, P., and Larionov, M.: First meteorite
  recovery based on observations by the Finnish Fireball Network, in:
  Proceedings of the International Meteor Conference, p. 162–169, Giron,
  France, 2014.

\bibitem[{{Hapgood} et~al.(2021){Hapgood}, {Angling}, {Attrill}, {Bisi},
  {Cannon}, {Dyer}, {Eastwood}, {Elvidge}, {Gibbs}, {Harrison}, {Hord},
  {Horne}, {Jackson}, {Jones}, {Machin}, {Mitchell}, {Preston}, {Rees},
  {Rogers}, {Routledge}, {Ryden}, {Tanner}, {Thomson}, {Wild}, and
  {Willis}}]{Hapgood2021}
{Hapgood}, M., {Angling}, M.~J., {Attrill}, G., {Bisi}, M., {Cannon}, P.~S.,
  {Dyer}, C., {Eastwood}, J.~P., {Elvidge}, S., {Gibbs}, M., {Harrison}, R.~A.,
  {Hord}, C., {Horne}, R.~B., {Jackson}, D.~R., {Jones}, B., {Machin}, S.,
  {Mitchell}, C.~N., {Preston}, J., {Rees}, J., {Rogers}, N.~C., {Routledge},
  G., {Ryden}, K., {Tanner}, R., {Thomson}, A. W.~P., {Wild}, J.~A., and
  {Willis}, M.: {Development of Space Weather Reasonable Worst Case Scenarios
  for the UK National Risk Assessment}, Space Weather, 19, e2020SW002\,593,
  \doi{10.1029/2020SW002593}, 2021.

\bibitem[{Hayakawa et~al.(2024)Hayakawa, Ebihara, Mishev, Koldobskiy, Kusano,
  Bechet, Yashiro, Iwai, Shinbori, Mursula, Miyake, Shiota, Silveira, Stuart,
  Oliveira, Akiyama, Ohnishi, and Miyoshi}]{Hayakawa2024}
Hayakawa, H., Ebihara, Y., Mishev, A., Koldobskiy, S., Kusano, K., Bechet, S.,
  Yashiro, S., Iwai, K., Shinbori, A., Mursula, K., Miyake, F., Shiota, D.,
  Silveira, M. V.~D., Stuart, R., Oliveira, D.~M., Akiyama, S., Ohnishi, K.,
  and Miyoshi, Y.: The Solar and Geomagnetic Storms in May 2024: A Flash Data
  Report, arXiv preprint, \doi{10.48550/arXiv.2407.07665}, 2024.

\bibitem[{{He} et~al.(2023){He}, {Yao}, {Ni}, {Cao}, {Ye}, {Guo}, {Li}, {Ren},
  {Yue}, {Zhang}, {Wei}, {Zhang}, and {Pu}}]{he2023}
{He}, F., {Yao}, Z., {Ni}, B., {Cao}, X., {Ye}, S., {Guo}, R., {Li}, J., {Ren},
  Z., {Yue}, X., {Zhang}, Y., {Wei}, Y., {Zhang}, X., and {Pu}, Z.: {Sawtooth
  and dune auroras simultaneously driven by waves around the plasmapause},
  Earth and Planetary Physics, 7, 237--246, \doi{10.26464/epp2023023}, 2023.

\bibitem[{Herlingshaw et~al.(2024)Herlingshaw, Lach, Dayton-Oxland, Bruus,
  Karvinen, Ledvina, Partamies, Grandin, Spijkers, Nishimura, Knudsen,
  Ladbrook, Martinis, Gallardo-Lacourt, Dyer, Mielke, Ratzlaff, Evans, Helin,
  Kuzub, Mathieu, Thomas, Glad, Donovan, Syrjäsuo, Cordon, Andersen, and
  Legg}]{ARCTICS_handbook}
Herlingshaw, K., Lach, D., Dayton-Oxland, R., Bruus, E., Karvinen, E., Ledvina,
  V., Partamies, N., Grandin, M., Spijkers, M., Nishimura, Y., Knudsen, D.,
  Ladbrook, L., Martinis, C., Gallardo-Lacourt, B., Dyer, A., Mielke, L.,
  Ratzlaff, C., Evans, L., Helin, M., Kuzub, J., Mathieu, B., Thomas, N., Glad,
  M., Donovan, E., Syrjäsuo, M., Cordon, S., Andersen, J., and Legg, C.:
  {ARCTICS Aurora Field Guide And Handbook For Citizen Science}, Zenodo,
  \doi{10.5281/zenodo.13932081}, 2024.

\bibitem[{{Johnsen}(2013)}]{johnsen2013}
{Johnsen}, M.~G.: {Real-time determination and monitoring of the auroral
  electrojet boundaries}, Journal of Space Weather and Space Climate, 3, A28,
  \doi{10.1051/swsc/2013050}, 2013.

\bibitem[{{Karttunen}(1921)}]{ursa2021}
{Karttunen}, H.: Ursan historia : Tähtitieteellinen yhdistys Ursa 1921–2021,
  Tähtitieteellinen Yhdistys Ursa Ry, ISBN 9789525985986, 1921.

\bibitem[{{Kataoka} et~al.(2024){Kataoka}, {Miyoshi}, {Shiokawa}, {Nishitani},
  {Keika}, {Amano}, and {Seki}}]{kataoka2024}
{Kataoka}, R., {Miyoshi}, Y., {Shiokawa}, K., {Nishitani}, N., {Keika}, K.,
  {Amano}, T., and {Seki}, K.: {Magnetic Storm-Time Red Aurora as Seen From
  Hokkaido, Japan on 1 December 2023 Associated With High-Density Solar Wind},
  Geophysical Research Letters, 51, e2024GL108\,778,
  \doi{10.1029/2024GL108778}, 2024.

\bibitem[{{Kosar} et~al.(2018{\natexlab{a}}){Kosar}, {MacDonald}, {Case}, and
  {Heavner}}]{cosar2018_data}
{Kosar}, B.~C., {MacDonald}, E.~A., {Case}, N.~A., and {Heavner}, M.:
  {Aurorasaurus Database of Real-Time, Crowd-Sourced Aurora Data for Space
  Weather Research}, Earth and Space Science, 5, 970--980,
  \doi{10.1029/2018EA000454}, 2018{\natexlab{a}}.

\bibitem[{{Kosar} et~al.(2018{\natexlab{b}}){Kosar}, {MacDonald}, {Case},
  {Zhang}, {Mitchell}, and {Viereck}}]{cosar2018_boundaries}
{Kosar}, B.~C., {MacDonald}, E.~A., {Case}, N.~A., {Zhang}, Y., {Mitchell},
  E.~J., and {Viereck}, R.: {A case study comparing citizen science aurora data
  with global auroral boundaries derived from satellite imagery and empirical
  models}, Journal of Atmospheric and Solar-Terrestrial Physics, 177, 274--282,
  \doi{10.1016/j.jastp.2018.05.006}, 2018{\natexlab{b}}.

\bibitem[{{Kwak} et~al.(2024){Kwak}, {Kim}, {Kim}, {Miyashita}, {Yang}, {Park},
  {Lim}, {Jung}, {Kam}, {Lee}, {Lee}, {Yoo}, {Lee}, {Kwon}, {Seough}, {Nam},
  {Lee}, {Hong}, {Sohn}, {Kwak}, {Kwak}, {Kim}, {Kim}, {Cho}, {Park}, {Lee},
  {Nguyen}, and {Talha}}]{kwak2024}
{Kwak}, Y.-S., {Kim}, J.-H., {Kim}, S., {Miyashita}, Y., {Yang}, T., {Park},
  S.-H., {Lim}, E.-K., {Jung}, J., {Kam}, H., {Lee}, J., {Lee}, H., {Yoo},
  J.-H., {Lee}, H., {Kwon}, R.-Y., {Seough}, J., {Nam}, U.-W., {Lee}, W.~K.,
  {Hong}, J., {Sohn}, J., {Kwak}, J., {Kwak}, H., {Kim}, R.-S., {Kim}, Y.-H.,
  {Cho}, K.-S., {Park}, J., {Lee}, J., {Nguyen}, H. N.~H., and {Talha}, M.:
  {Observational Overview of the May 2024 G5-Level Geomagnetic Storm: From
  Solar Eruptions to Terrestrial Consequences}, Journal of Astronomy and Space
  Sciences, 41, 171--194, \doi{10.5140/JASS.2024.41.3.171}, 2024.

\bibitem[{{Ledvina} et~al.(2023){Ledvina}, {Brandt}, {MacDonald}, {Frissell},
  {Anderson}, {Chen}, {French}, {Di Mare}, {Grover}, {Battams}, {Sigsbee},
  {Gallardo-Lacourt}, {Lach}, {Shaw}, {Hunnekuhl}, {Kosar}, {Barkhouse},
  {Young}, {Kedhambadi}, {Ozturk}, {Claudepierre}, {Dong}, {Witteman}, {Kuzub},
  and {Sinha}}]{ledvina2023}
{Ledvina}, V., {Brandt}, L., {MacDonald}, E., {Frissell}, N., {Anderson}, J.,
  {Chen}, T.~Y., {French}, R.~J., {Di Mare}, F., {Grover}, A., {Battams}, K.,
  {Sigsbee}, K., {Gallardo-Lacourt}, B., {Lach}, D., {Shaw}, J.~A.,
  {Hunnekuhl}, M., {Kosar}, B., {Barkhouse}, W., {Young}, T., {Kedhambadi}, C.,
  {Ozturk}, D.~S., {Claudepierre}, S.~G., {Dong}, C., {Witteman}, A., {Kuzub},
  J., and {Sinha}, G.: {Agile collaboration: Citizen science as a
  transdisciplinary approach to heliophysics}, Frontiers in Astronomy and Space
  Sciences, 10, 1165\,254, \doi{10.3389/fspas.2023.1165254}, 2023.

\bibitem[{Liang et~al.(2024)Liang, Gillies, Spanswick, and
  Donovan}]{Liang_2023}
Liang, J., Gillies, D.~M., Spanswick, E., and Donovan, E.~F.: Converting
  TREx-RGB green-channel data to 557.7 nm auroral intensity: Methodology and
  initial results, Earth and Planetary Physics, 8, 258--274,
  \doi{10.26464/epp2023063}, 2024.

\bibitem[{{Lilensten} and {Blelly}(2002)}]{lilensten2002}
{Lilensten}, J. and {Blelly}, P.~L.: {The TEC and F2 parameters as tracers of
  the ionosphere and thermosphere}, Journal of Atmospheric and
  Solar-Terrestrial Physics, 64, 775--793, \doi{10.1016/S1364-6826(02)00079-2},
  2002.

\bibitem[{{Lu} et~al.(2016){Lu}, {Richmond}, {L{\"u}hr}, and {Paxton}}]{lu2016}
{Lu}, G., {Richmond}, A.~D., {L{\"u}hr}, H., and {Paxton}, L.: {High-latitude
  energy input and its impact on the thermosphere}, Journal of Geophysical
  Research (Space Physics), 121, 7108--7124, \doi{10.1002/2015JA022294}, 2016.

\bibitem[{{Lummerzheim} and {Lilensten}(1994)}]{lummerzheim1994}
{Lummerzheim}, D. and {Lilensten}, J.: {Electron transport and energy
  degradation in the ionosphere: Evaluation of the numerical solution,
  comparison with laboratory experiments and auroral observations}, Annales
  Geophysicae, 12, 1039--1051, \doi{10.1007/s00585-994-1039-7}, 1994.

\bibitem[{MacDonald et~al.(2015)MacDonald, Case, Clayton, Hall, Heavner,
  Lalone, Patel, and Tapia}]{macdonald2015}
MacDonald, E.~A., Case, N.~A., Clayton, J.~H., Hall, M.~K., Heavner, M.,
  Lalone, N., Patel, K.~G., and Tapia, A.: Aurorasaurus: A citizen science
  platform for viewing and reporting the aurora, Space Weather, 13, 548--559,
  \doi{https://doi.org/10.1002/2015SW001214}, 2015.

\bibitem[{{MacDonald} et~al.(2018){MacDonald}, {Donovan}, {Nishimura}, {Case},
  {Gillies}, {Gallardo-Lacourt}, {Archer}, {Spanswick}, {Bourassa}, {Connors},
  {Heavner}, {Jackel}, {Kosar}, {Knudsen}, {Ratzlaff}, and
  {Schofield}}]{macdonald2018}
{MacDonald}, E.~A., {Donovan}, E., {Nishimura}, Y., {Case}, N.~A., {Gillies},
  D.~M., {Gallardo-Lacourt}, B., {Archer}, W.~E., {Spanswick}, E.~L.,
  {Bourassa}, N., {Connors}, M., {Heavner}, M., {Jackel}, B., {Kosar}, B.,
  {Knudsen}, D.~J., {Ratzlaff}, C., and {Schofield}, I.: {New science in plain
  sight: Citizen scientists lead to the discovery of optical structure in the
  upper atmosphere}, Science Advances, 4, eaaq0030,
  \doi{10.1126/sciadv.aaq0030}, 2018.

\bibitem[{{Malacara}(2011)}]{malacara2011}
{Malacara}, D.: {Color Vision and Colorimetry, Theory and Applications}, SPIE
  Press, Bellingham, WA, USA, ISBN 0-8194-4228-3, 2011.

\bibitem[{Martinis et~al.(2021)Martinis, Nishimura, Wroten, Bhatt, Dyer,
  Baumgardner, and Gallardo-Lacourt}]{Martinis2021}
Martinis, C., Nishimura, Y., Wroten, J., Bhatt, A., Dyer, A., Baumgardner, J.,
  and Gallardo-Lacourt, B.: First Simultaneous Observation of STEVE and SAR Arc
  Combining Data From Citizen Scientists, 630.0 nm All-Sky Images, and
  Satellites, Geophysical Research Letters, 48, e2020GL092\,169,
  \doi{https://doi.org/10.1029/2020GL092169}, 2021.

\bibitem[{{Martinis} et~al.(2022){Martinis}, {Griffin}, {Gallardo-Lacourt},
  {Wroten}, {Nishimura}, {Baumgardner}, and {Knudsen}}]{martinis2022}
{Martinis}, C., {Griffin}, I., {Gallardo-Lacourt}, B., {Wroten}, J.,
  {Nishimura}, Y., {Baumgardner}, J., and {Knudsen}, D.~J.: {Rainbow of the
  Night: First Direct Observation of a SAR Arc Evolving Into STEVE},
  Geophysical Research Letters, 49, e98\,511, \doi{10.1029/2022GL098511}, 2022.

\bibitem[{Mende and Turner(2019)}]{Mende2019}
Mende, S.~B. and Turner, C.: Color Ratios of Subauroral (STEVE) Arcs, Journal
  of Geophysical Research: Space Physics, 124, 5945--5955,
  \doi{https://doi.org/10.1029/2019JA026851}, 2019.

\bibitem[{Moilanen and Gritsevich(2022)}]{moilanen2022}
Moilanen, J. and Gritsevich, M.: {Light scattering by airborne ice crystals –
  An inventory of atmospheric halos}, Journal of Quantitative Spectroscopy and
  Radiative Transfer, 290, \doi{10.1016/j.jqsrt.2022.10831}, 2022.

\bibitem[{{Newell} and {Gjerloev}(2011)}]{SuperMAG_electrojet}
{Newell}, P.~T. and {Gjerloev}, J.~W.: {Evaluation of SuperMAG auroral
  electrojet indices as indicators of substorms and auroral power}, Journal of
  Geophysical Research (Space Physics), 116, A12\,211,
  \doi{10.1029/2011JA016779}, 2011.

\bibitem[{{Newell} and {Gjerloev}(2012)}]{SuperMAG_ringcurrent}
{Newell}, P.~T. and {Gjerloev}, J.~W.: {SuperMAG-based partial ring current
  indices}, Journal of Geophysical Research (Space Physics), 117, A05\,215,
  \doi{10.1029/2012JA017586}, 2012.

\bibitem[{{Newell} et~al.(2014){Newell}, {Liou}, {Zhang}, {Sotirelis},
  {Paxton}, and {Mitchell}}]{newell2014}
{Newell}, P.~T., {Liou}, K., {Zhang}, Y., {Sotirelis}, T., {Paxton}, L.~J., and
  {Mitchell}, E.~J.: {OVATION Prime-2013: Extension of auroral precipitation
  model to higher disturbance levels}, Space Weather, 12, 368--379,
  \doi{10.1002/2014SW001056}, 2014.

\bibitem[{{Nishimura} et~al.(2022){Nishimura}, Bruus, Karvinen, Martinis, Dyer,
  Kangas, Rikala, Donovan, Nishitani, and Ruohoniemi}]{nishimura2022}
{Nishimura}, Y., Bruus, E., Karvinen, E., Martinis, C.~R., Dyer, A., Kangas,
  L., Rikala, H.~K., Donovan, E.~F., Nishitani, N., and Ruohoniemi, J.~M.:
  {Interaction Between Proton Aurora and Stable Auroral Red Arcs Unveiled by
  Citizen Scientist Photographs}, Journal of Geophysical Research (Space
  Physics), 127, \doi{10.1029/2022JA030570}, 2022.

\bibitem[{{Nishimura} et~al.(2023){Nishimura}, {Dyer}, {Kangas}, {Donovan}, and
  {Angelopoulos}}]{nishimura2023}
{Nishimura}, Y., {Dyer}, A., {Kangas}, L., {Donovan}, E., and {Angelopoulos},
  V.: {Unsolved problems in Strong Thermal Emission Velocity Enhancement
  (STEVE) and the picket fence}, Frontiers in Astronomy and Space Sciences, 10,
  3, \doi{10.3389/fspas.2023.1087974}, 2023.

\bibitem[{{Nishitani} et~al.(2019){Nishitani}, {Ruohoniemi}, {Lester}, {Baker},
  {Koustov}, {Shepherd}, {Chisham}, {Hori}, {Thomas}, {Makarevich},
  {Marchaudon}, {Ponomarenko}, {Wild}, {Milan}, {Bristow}, {Devlin}, {Miller},
  {Greenwald}, {Ogawa}, and {Kikuchi}}]{Nishitani2019}
{Nishitani}, N., {Ruohoniemi}, J.~M., {Lester}, M., {Baker}, J. B.~H.,
  {Koustov}, A.~V., {Shepherd}, S.~G., {Chisham}, G., {Hori}, T., {Thomas},
  E.~G., {Makarevich}, R.~A., {Marchaudon}, A., {Ponomarenko}, P., {Wild},
  J.~A., {Milan}, S.~E., {Bristow}, W.~A., {Devlin}, J., {Miller}, E.,
  {Greenwald}, R.~A., {Ogawa}, T., and {Kikuchi}, T.: {Review of the
  accomplishments of mid-latitude Super Dual Auroral Radar Network (SuperDARN)
  HF radars}, Progress in Earth and Planetary Science, 6, 27,
  \doi{10.1186/s40645-019-0270-5}, 2019.

\bibitem[{{Oliveira} et~al.(2021){Oliveira}, {Zesta}, {Mehta}, {Licata},
  {Pilinski}, {Tobiska}, and {Hayakawa}}]{Oliveira2021}
{Oliveira}, D.~M., {Zesta}, E., {Mehta}, P.~M., {Licata}, R.~J., {Pilinski},
  M.~D., {Tobiska}, W.~K., and {Hayakawa}, H.: The Current State and Future
  Directions of Modeling Thermosphere Density Enhancements During Extreme
  Magnetic Storms, Frontiers in Astronomy and Space Sciences, 8, 764\,144,
  \doi{10.3389/fspas.2021.764144}, 2021.

\bibitem[{{Oughton} et~al.(2017){Oughton}, {Skelton}, {Horne}, {Thomson}, and
  {Gaunt}}]{oughton2017}
{Oughton}, E.~J., {Skelton}, A., {Horne}, R.~B., {Thomson}, A. W.~P., and
  {Gaunt}, C.~T.: {Quantifying the daily economic impact of extreme space
  weather due to failure in electricity transmission infrastructure}, Space
  Weather, 15, 65--83, \doi{10.1002/2016SW001491}, 2017.

\bibitem[{{Palmroth} et~al.(2019){Palmroth}, {Grandin}, {Helin}, {Koski},
  {Oksanen}, {Glad}, {Valonen}, {Saari}, {Bruus}, {Norberg}, {Viljanen},
  {Kauristie}, and {Verronen}}]{palmroth2019}
{Palmroth}, M., {Grandin}, M., {Helin}, M., {Koski}, P., {Oksanen}, A., {Glad},
  M.~A., {Valonen}, R., {Saari}, K., {Bruus}, E., {Norberg}, J., {Viljanen},
  A., {Kauristie}, K., and {Verronen}, P.~T.: {Citizen Scientists Discover a
  New Auroral Form: Dunes Provide Insight Into the Upper Atmosphere}, AGU
  Advances, 1, {e00\,133}, \doi{10.1029/2019AV000133}, 2019.

\bibitem[{{Palmroth} et~al.(2021){Palmroth}, {Grandin}, {Sarris}, {Doornbos},
  {Tourgaidis}, {Aikio}, {Buchert}, {Clilverd}, {Dandouras}, {Heelis},
  {Hoffmann}, {Ivchenko}, {Kervalishvili}, {Knudsen}, {Kotova}, {Liu},
  {Malaspina}, {March}, {Marchaudon}, {Marghitu}, {Matsuo}, {Miloch},
  {Moretto-J{\o}rgensen}, {Mpaloukidis}, {Olsen}, {Papadakis}, {Pfaff},
  {Pirnaris}, {Siemes}, {Stolle}, {Suni}, {van den IJssel}, {Verronen},
  {Visser}, and {Yamauchi}}]{palmroth2021}
{Palmroth}, M., {Grandin}, M., {Sarris}, T., {Doornbos}, E., {Tourgaidis}, S.,
  {Aikio}, A., {Buchert}, S., {Clilverd}, M.~A., {Dandouras}, I., {Heelis}, R.,
  {Hoffmann}, A., {Ivchenko}, N., {Kervalishvili}, G., {Knudsen}, D.~J.,
  {Kotova}, A., {Liu}, H.-L., {Malaspina}, D.~M., {March}, G., {Marchaudon},
  A., {Marghitu}, O., {Matsuo}, T., {Miloch}, W.~J., {Moretto-J{\o}rgensen},
  T., {Mpaloukidis}, D., {Olsen}, N., {Papadakis}, K., {Pfaff}, R., {Pirnaris},
  P., {Siemes}, C., {Stolle}, C., {Suni}, J., {van den IJssel}, J., {Verronen},
  P.~T., {Visser}, P., and {Yamauchi}, M.: {Lower-thermosphere-ionosphere (LTI)
  quantities: current status of measuring techniques and models}, Annales
  Geophysicae, 39, 189--237, \doi{10.5194/angeo-39-189-2021}, 2021.

\bibitem[{Papitashvili and King(2020)}]{OMNI_data_1hr}
Papitashvili, N.~E. and King, J.~H.: {OMNI Hourly Data [Data set]}, NASA Space
  Physics Data Facility, \doi{10.48322/1shr-ht18}, 2020.

\bibitem[{{Parker} and {Linares}(2024)}]{parker2024}
{Parker}, W.~E. and {Linares}, R.: {Satellite Drag Analysis During the May 2024
  Gannon Geomagnetic Storm}, Journal of Spacecraft and Rockets, 61, 1412--1416,
  \doi{10.2514/1.A36164}, 2024.

\bibitem[{{Picone} et~al.(2002){Picone}, {Hedin}, {Drob}, and
  {Aikin}}]{picone2002}
{Picone}, J.~M., {Hedin}, A.~E., {Drob}, D.~P., and {Aikin}, A.~C.:
  {NRLMSISE-00 empirical model of the atmosphere: Statistical comparisons and
  scientific issues}, Journal of Geophysical Research (Space Physics), 107,
  1468, \doi{10.1029/2002JA009430}, 2002.

\bibitem[{{Robert} et~al.(2023){Robert}, {Barthelemy}, {Cessateur},
  {Woelffl{\'e}}, {Lamy}, {Bouriat}, {Gullikstad Johnsen},
  {Br{\"a}ndstr{\"o}m}, and {Biree}}]{robert2023}
{Robert}, E., {Barthelemy}, M., {Cessateur}, G., {Woelffl{\'e}}, A., {Lamy},
  H., {Bouriat}, S., {Gullikstad Johnsen}, M., {Br{\"a}ndstr{\"o}m}, U., and
  {Biree}, L.: {Reconstruction of electron precipitation spectra at the top of
  the upper atmosphere using 427.8 nm auroral images}, Journal of Space Weather
  and Space Climate, 13, 30, \doi{10.1051/swsc/2023028}, 2023.

\bibitem[{{Sandholt} et~al.(2002){Sandholt}, {Carlson}, and
  Egeland}]{Sandholt2002}
{Sandholt}, P.~E., {Carlson}, H.~C., and Egeland, A.: {Optical Aurora}, pp.
  33--51, Springer, Dordrecht, Netherlands, ISBN 978-0-306-47969-4,
  \doi{10.1007/0-306-47969-9_3}, 2002.

\bibitem[{Schnapf et~al.(1987)Schnapf, Kraft, and Baylor}]{schnapf1987}
Schnapf, J., Kraft, T., and Baylor, D.: Spectral sensitivity of human cone
  photoreceptors, Nature, 325, 439--441, \doi{10.1038/325439a0}, 1987.

\bibitem[{{Semeter} et~al.(2020){Semeter}, {Hunnekuhl}, {MacDonald}, {Hirsch},
  {Zeller}, {Chernenkoff}, and {Wang}}]{semeter2020}
{Semeter}, J., {Hunnekuhl}, M., {MacDonald}, E., {Hirsch}, M., {Zeller}, N.,
  {Chernenkoff}, A., and {Wang}, J.: {The Mysterious Green Streaks Below
  STEVE}, AGU Advances, 1, e00\,183,
  \doi{10.1029/2020AV00018310.1002/essoar.10502878.2}, 2020.

\bibitem[{{Shepherd}(2014)}]{shepherd2014}
{Shepherd}, S.~G.: {Altitude-adjusted corrected geomagnetic coordinates:
  Definition and functional approximations}, Journal of Geophysical Research
  (Space Physics), 119, 7501--7521, \doi{10.1002/2014JA020264}, 2014.

\bibitem[{Shiokawa et~al.(2019)Shiokawa, Otsuka, and Connors}]{shiokawa2019}
Shiokawa, K., Otsuka, Y., and Connors, M.: Statistical Study of
  Auroral/Resonant-Scattering 427.8-nm Emission Observed at Subauroral
  Latitudes Over 14 Years, Journal of Geophysical Research: Space Physics, 124,
  9293--9301, \doi{https://doi.org/10.1029/2019JA026704}, 2019.

\bibitem[{{Sigernes} et~al.(2011){Sigernes}, {Dyrland}, {Brekke}, {Chernouss},
  {Lorentzen}, {Oksavik}, and {Deehr}}]{sigernes2011}
{Sigernes}, F., {Dyrland}, M., {Brekke}, P., {Chernouss}, S., {Lorentzen},
  D.~A., {Oksavik}, K., and {Deehr}, C.~S.: Two methods to forecast auroral
  displays, Journal of Space Weather and Space Climate, 1, A03,
  \doi{10.1051/swsc/2011003}, 2011.

\bibitem[{{Siscoe} et~al.(2006){Siscoe}, {Crooker}, and {Clauer}}]{siscoe2006}
{Siscoe}, G., {Crooker}, N.~U., and {Clauer}, C.~R.: {Dst of the Carrington
  storm of 1859}, Advances in Space Research, 38, 173--179,
  \doi{10.1016/j.asr.2005.02.102}, 2006.

\bibitem[{Spogli et~al.(2024)Spogli, Alberti, Bagiacchi, Cafarella, Cesaroni,
  Cianchini, Coco, Di~Mauro, Ghidoni, Giannattasio, Ippolito, Marcocci,
  Pezzopane, Pica, Pignalberi, Perrone, Romano, Sabbagh, Scotto, Spadoni,
  Tozzi, and Viola}]{spogli2024}
Spogli, L., Alberti, T., Bagiacchi, P., Cafarella, L., Cesaroni, C., Cianchini,
  G., Coco, I., Di~Mauro, D., Ghidoni, R., Giannattasio, F., Ippolito, A.,
  Marcocci, C., Pezzopane, M., Pica, E., Pignalberi, A., Perrone, L., Romano,
  V., Sabbagh, D., Scotto, C., Spadoni, S., Tozzi, R., and Viola, M.: {The
  effects of the May 2024 Mother’s Day superstorm over the Mediterranean
  sector: from data to public communication}, Annals of Geophysics, 67, PA218,
  \doi{10.4401/ag-9117}, 2024.

\bibitem[{{Usoskin} et~al.(2023){Usoskin}, {Miyake}, {Baroni}, {Brehm},
  {Dalla}, {Hayakawa}, {Hudson}, {Jull}, {Knipp}, {Koldobskiy}, {Maehara},
  {Mekhaldi}, {Notsu}, {Poluianov}, {Rozanov}, {Shapiro}, {Spiegl},
  {Sukhodolov}, {Uusitalo}, and {Wacker}}]{Usoskin2023}
{Usoskin}, I., {Miyake}, F., {Baroni}, M., {Brehm}, N., {Dalla}, S.,
  {Hayakawa}, H., {Hudson}, H., {Jull}, A.~J.~T., {Knipp}, D., {Koldobskiy},
  S., {Maehara}, H., {Mekhaldi}, F., {Notsu}, Y., {Poluianov}, S., {Rozanov},
  E., {Shapiro}, A., {Spiegl}, T., {Sukhodolov}, T., {Uusitalo}, J., and
  {Wacker}, L.: {Extreme Solar Events: Setting up a Paradigm}, Space Science
  Reviews, 219, 73, \doi{10.1007/s11214-023-01018-1}, 2023.

\bibitem[{{Whiter} et~al.(2021){Whiter}, {Sundberg}, {Lanchester}, {Dreyer},
  {Partamies}, {Ivchenko}, {Zaccaria Di Fraia}, {Oliver}, {Serpell-Stevens},
  {Shaw-Diaz}, and {Braunersreuther}}]{whiter2021}
{Whiter}, D.~K., {Sundberg}, H., {Lanchester}, B.~S., {Dreyer}, J.,
  {Partamies}, N., {Ivchenko}, N., {Zaccaria Di Fraia}, M., {Oliver}, R.,
  {Serpell-Stevens}, A., {Shaw-Diaz}, T., and {Braunersreuther}, T.:
  {Fine-scale dynamics of fragmented aurora-like emissions}, Annales
  Geophysicae, 39, 975--989, \doi{10.5194/angeo-39-975-2021}, 2021.

\bibitem[{{Whiter} et~al.(2023){Whiter}, {Partamies}, {Gustavsson}, and
  {Kauristie}}]{whiter2023}
{Whiter}, D.~K., {Partamies}, N., {Gustavsson}, B., and {Kauristie}, K.: {The
  altitude of green OI 557.7 nm and blue N$_{2}$$^{+}$ 427.8 nm aurora},
  Annales Geophysicae, 41, 1--12, \doi{10.5194/angeo-41-1-2023}, 2023.

\end{thebibliography}
\end{document}